\documentclass{aa}
\usepackage{graphics}
\usepackage{supertabular_aa}
\begin{document}

\thesaurus{20(08.09.2 HD 93521; 09.12.1; 09.13.2; 13.21.3; 13.21.5)}

\title{The ORFEUS\,II  Echelle Spectrum of HD\,93521:\\
A reference for interstellar molecular hydrogen}
\author{
J.\,Barnstedt\inst{1} \and
W.\,Gringel\inst{1} \and
N.\,Kappelmann\inst{1} \and
M.\,Grewing\inst{1,2}
}

\offprints{J. Barnstedt\\
e-mail: barnstedt@astro.uni-tuebingen.de}

\institute{Institut f\"{u}r Astronomie und Astrophysik, Abt. Astronomie, Eberhard-Karls-%
Universit\"{a}t T\"{u}bingen, Waldh\"{a}userstr. 64, D-72076 T\"{u}bingen, Germany \and
Institut de Radio Astronomie Millim\'{e}trique (IRAM), 300 Rue de la Piscine, F-38406 Saint
Martin d'H\`{e}res, France}

\titlerunning{The {\it ORFEUS\,II}  Echelle Spectrum of HD\,93521}

\date{Received December 13, 1999; accepted January 21, 2000 }
%\date{accepted}

\maketitle

\begin{abstract}

During the second flight of the {\it ORFEUS-SPAS} mission in
November/December 1996, the Echelle spectrometer was used extensively
by the Principal and Guest Investigator teams as one of the two focal
plane instruments of the {\it ORFEUS} telescope. The spectrum of
\object{HD\,93521} was obtained during this mission with a total
integration time of 1740\,s. This spectrum shows numerous sharp
interstellar absorption lines. We identified 198 lines of molecular
hydrogen including at least 7 lines with a high velocity component.
Also most of the 67 identified interstellar metal lines are visible
with a high velocity component.

We present plots of the complete {\it ORFEUS\,II} Echelle spectrum
together with tables of all identified interstellar absorption lines
including all 14 detectable H\,I lines. In addition several identified
stellar lines, partially with narrow absorption components, and stellar
wind lines are given in a separate table.

\end{abstract}

\section{Introduction}

The star HD\,93521 is a high galactic latitude O-star. It has been used
for many years as a tracer of the interstellar gas in the galactic
halo, for which it is well suited due to its brightness ($V=7.04$),
high galactic latitude ($l\,=\,183\degr$, $b\,=\,62\degr$) and large
rotational velocity ($v \sin i \approx 400$\,km\,s$^{-1}$, Lennon et
al. \cite{lennon}). Together with a $z$-distance of about 1.5\,kpc
(Irvine \cite{irvine}), HD\,93521 is an ideal candidate for studying
kinematics of the halo gas, since galactic rotation effects should be
very small (Spitzer \& Fitzpatrick \cite{spitzer1}).

With the Echelle spectrometer of the {\it ORFEUS} telescope (Orbiting
and Retrievable Far and Extreme Ultraviolet Spectrometer) it was for
the first time possible to observe all available absorption lines of
interstellar molecular hydrogen towards HD\,93521. In the gathered
spectra we identified 198 H$_{2}$-lines. These lines are very narrow
and strong but unsaturated, which was the reason why many of them were
used for wavelength calibration of the {\it ORFEUS} Echelle
spectrometer. Also 67 other interstellar absorption lines and 14 lines
of the Lyman series were identified. For completeness stellar and
stellar wind absorption lines are shown too.

We present the whole Echelle spectrum of HD\,93521, obtained during the
second {\it ORFEUS-SPAS} mission in November/December 1996. This
spectrum has a good signal to noise ratio and the spectral resolution
achieved is somewhat higher than the claimed resolution of
$\lambda$/$\Delta$$\lambda$\,=\,10.000 (Barnstedt et al.
\cite{barnstedt}).

The plots presented in the appendix show one Echelle order per plot for
wavelengths above 1130\,{\AA} (Echelle orders 40 to 49), and half an
Echelle order per plot for Echelle orders 50 to 61
($\lambda$\,$<$\,1130\,{\AA}) where all H$_{2}$-lines are included.

\section{Data reduction and line identification}

Two separate observations of HD\,93521 were obtained during two
successive orbits with a total integration time of 1740\,s (ORFEUS
observation IDs 2276\_2 and 2276\_3, observation date: day 333 of 1996,
GMT 04:56:05 -- 05:14:05 and GMT 06:28:05 -- 06:39:05). The two echelle
images were coadded and then the standard extraction procedure was
applied (Barnstedt et al. \cite{barnstedt}) without any smoothing. An
additional radial velocity correction of $-10$\,km\,s$^{-1}$ was
applied, which corrects the wavelength scale for the fact that the star
was not absolutely centered in the entrance diaphragm of the telescope.
The maximum uncertainty due to the 20{\arcsec} diameter of the
diaphragm was $\pm$$36$\,km\,s$^{-1}$, so the deviation of
$-10$\,km\,s$^{-1}$ corresponds to a pointing offset of 3{\arcsec},
which is an excellent value for the {\it ASTRO-SPAS} satellite. The
value of $-10$\,km\,s$^{-1}$ was estimated by comparing the observed
radial velocity components with those already published (Spitzer \&
Fitzpatrick \cite{spitzer1}). The wavelength scale is heliocentric, a
LSR scale would require an additional correction of
$-1.6$\,km\,s$^{-1}$, which is negligible.

As with all echelle spectra, the signal to noise ratio is best in the
centre of the echelle orders, while it is reduced by a factor of about
1.4 at both ends of each order. Due to the blaze curve being not fully
centered, the signal to noise ratio at the short wavelength end is
significantly better than at the long wavelength end of each echelle
order. There is also a slight deviation in the wavelength calibration
at the short wavelength end of each order, which affects a wavelength
range of about 5\% in each order.

For line identifications we used the following line catalogues:
\begin{enumerate}
\item H$_{2}$-lines:
\begin{itemize}
\item[-] Morton \& Dinerstein (\cite{morton1}), except L10P1
\item[-] Abgrall et al. (\cite{abgrall}), correct wavelength for L10P1: 982.834\,{\AA}
\end{itemize}
\item Metal lines and H\,I:
\begin{itemize}
\item[-] Morton (\cite{morton2})
\item[-] Kurucz CD No. 23, web-page (Kurucz \cite{kurucz})
\item[-] Kelly (\cite{kelly}; and web-page)
\item[-] Feibelman \& Johannson (\cite{feibelman})
\end{itemize}
\end{enumerate}

Identified lines from IUE spectra of HD\,93521 above 1170\,{\AA} are
listed by Ramella et al. (\cite{ramella}), with exception of two
interstellar lines: $\lambda$1260.4, which is Si\,II and not Si\,III,
and $\lambda$1304.4, which is also Si\,II and not O\,I.

The tables and plots show blended lines also, for which a
non-ambiguous identification or estimate of the intensity is not
possible.

Lines with a lower energy level greater than zero are marked with an
asterisk, *. All stellar lines are marked with a bracketed asterisk,
(*), and stellar wind lines are marked as (w) in the plots as well as
in the tables.

\section{Discussion of spectral features}

The interstellar, stellar and wind absorption lines are visible in
several or different radial velocity components. We therefore list and
describe all occuring radial velocity components in Table\,1.

We present tables of identified interstellar and stellar absorption
lines. These tables show a running number for identification of the
lines in the plots shown in the appendix, the vacuum wavelength, the
$\log(gf)$-value, the number of the radial velocity component (VC)
applied as given in Table\,1, and some remarks or the transition for
the H$_{2}$-lines. We will present and discuss the tables of molecular
hydrogen lines, other interstellar lines, Lyman series lines and
stellar absorption lines.

\subsection{Radial velocity components}

Table\,1 lists 7 components of radial velocities used to identify
absorption lines and features in the spectrum. The first two
components are the interstellar absorptions at $-12$\,km\,s$^{-1}$ and
$-60$\,km\,s$^{-1}$, which are the two strongest of well known
interstellar components (Grewing et al. \cite {grewing}; Keenan et al.
\cite{keenan}; Spitzer \& Fitzpatrick \cite{spitzer1},
\cite{spitzer2}). No.\,3 gives the published value of the radial
velocity of HD\,93521 of $-16$\,km\,s$^{-1 }$({\it SIMBAD}). No.\,4 is the
velocity of the emission of the geocoronal Ly-$\alpha$ line. This
emission line results from a completely illuminated entrance aperture
of the Echelle spectrometer which had a projected diameter of
20{\arcsec}. The velocity of 36.5\,km\,s$^{-1}$ is the negative sum of
two wavelength corrections applied to this spectrum: the heliocentric
correction and the decentering correction (26.5\,km\,s$^{-1}$$ +
10$\,km\,s$^{-1}$).

Some stellar absorption lines show narrow absorption components
resulting from winds, which have been observed previously (Bjorkman et
al. \cite{bjorkman}), but which are varying in time. We have identified
two such components in serveral lines and they are listed as numbers 5
and 6 in Table\,1. Component 7 represents the radial velocities of the
strong Si\,III $\lambda$$\lambda$1300 triplets, which are also due to
stellar wind absorption (Massa \cite{massa}).

\begin{table}[ht]
\caption[]{Table of radial velocity components (VC).}

\begin{tabular*}{\hsize}{rrl}
\hline
VC  & Rad.vel. & Description \\
No. & [km/s]   & \\
\hline
1&  $-12$  & first (main) interstellar component \\
2&  $-60$  & high velocity interstellar component \\
3&  $-16$  & radial velocity of HD\,93521 \\
4& $36.5$ & geocoronal Ly-$\alpha$ emission\\
5& $-270$  & 1. wind feature in stellar absorption lines \\
6& $-340$  & 2. wind feature in stellar absorption lines \\
7& $-80$   & wind feature in Si\,III triplets \\
\hline
\end{tabular*}
\end{table}

\subsection{Interstellar molecular hydrogen}

For most of the H$_{2}$-lines only the main velocity component no.\,1
was observed, but for some unblended lines the high velocity component
could be seen also. A detailed discussion of column densities and curve
of growths will be published in a separate paper (Gringel et al., in
preparation).

Previous {\it Copernicus} measurements of selected H$_{2}$-lines only
led to an upper limit of $\log N(H_{2}) < 18.54$ (Savage et al.
\cite{savage}). The high velocity components were not detected by {\it
Copernicus}.

\paragraph{}

{\small

\tablecaption{Table of identified or possible interstellar molecular
hydrogen lines. VC is the velocity component as given in Table\,1.}

\tablehead{
  \hline
  No. & $\lambda$ [{\AA}] & $\log(f)$ & VC & Transition & Remarks \\
  \hline
}

\tabletail{
  \hline
  \multicolumn{5}{l}{Table 2, continued ...} \\
  \hline
}

\tablelasttail{
  \hline
}

\begin{supertabular}{rrrlll}
      1 &   918.411 & $-2.795$ & 1 &  L18P1 & blend \\
%article version only:
\shrinkheight{-0.5cm}
%referee version only:
%\shrinkheight{-1.5cm}
      2 &   918.427 & $-1.889$ & 1 &  W5Q3 & blend \\
      3 &   919.410 & $-2.430$ & 1 &  L18R2 & blend \\
      4 &   919.545 & $-2.345$ & 1 &  W5P3 & blend \\
      5 &   920.242 & $-2.767$ & 1 &  L18P2 \\
%article only:
\shrinkheight{-2cm}
      6 &   924.643 & $-2.412$ & 1 &  L17R1 \\
      7 &   925.173 & $-2.707$ & 1 &  L17P1 \\
      8 &   927.020 & $-2.628$ & 1 &  L17P2 \\
      9 &   928.437 & $-2.548$ & 1 &  L17R3 \\
     10 &   929.534 & $-1.470$ & 1 &  W4R0 & blend \\
     11 &   929.687 & $-1.810$ & 1 &  W4R1 & blend \\
     12 &   929.688 & $-2.595$ & 1 &  L17P3 & blend \\
     13 &   931.063 & $-1.991$ & 1 &  L16R0 \\
     14 &   931.732 & $-2.151$ & 1 &  L16R1 & blend \\
     15 &   931.779 & $-1.714$ & 1 &  W4Q2 & blend \\
     16 &   931.811 & $-1.979$ & 1 &  W4R3 & blend \\
     17 &   932.270 & $-2.621$ & 1 &  L16P1 \\
     18 &   932.606 & $-2.318$ & 1 &  W4P2 \\
     19 &   933.185 & $-2.563$ & 1 &  L17P4 & blend \\
     20 &   933.243 & $-2.202$ & 1 &  L16R2 & blend \\
     21 &   933.581 & $-1.714$ & 1 &  W4Q3 \\
     22 &   934.146 & $-2.643$ & 1 &  L16P2 \\
     23 &   934.789 & $-2.145$ & 1 &  W4P3 \\
     24 &   935.537 & $-2.498$ & 1 &  L17R5 & blend \\
     25 &   935.578 & $-2.236$ & 1 &  L16R3 & blend \\
     26 &   935.959 & $-1.717$ & 1 &  W4Q4 \\
     27 &   936.859 & $-2.679$ & 1 &  L16P3 \\
     28 &   938.468 & $-2.036$ & 1 &  L15R0 \\
     29 &   939.124 & $-2.207$ & 1 &  L15R1 \\
     30 &   939.710 & $-2.536$ & 1 &  L15P1 \\
     31 &   940.627 & $-2.256$ & 1 &  L15R2 \\
     32 &   941.601 & $-2.471$ & 1 &  L15P2 \\
     33 &   942.966 & $-2.293$ & 1 &  L15R3 \\
     34 &   944.331 & $-2.451$ & 1 &  L15P3 \\
     35 &   946.129 & $-2.342$ & 1 &  L15R4 & blend \\
     36 &   946.170 & $-2.958$ & 1 &  L14R0 & blend \\
     37 &   946.386 & $-1.889$ & 1 &  W3R1 & blend \\
%referee version only:
%\shrinkheight{-1.5cm}
     38 &   946.425 & $-1.207$ & 1 &  W3R0 & blend \\
     39 &   946.986 & $-2.728$ & 1 &  L14R1 & blend \\
     40 &   947.113 & $-1.876$ & 1 &  W3R2 \\
     41 &   947.425 & $-1.564$ & 1 &  W3Q1 \\
     42 &   947.517 & $-2.454$ & 1 &  L14P1 \\
     43 &   948.418 & $-1.893$ & 1 &  W3R3 & blend \\
     44 &   948.472 & $-1.917$ & 1 &  L14R2 & blend \\
     45 &   948.618 & $-1.564$ & 1 &  W3Q2 & blend \\
     46 &   949.355 & $-2.041$ & 1 &  L14P2 & blend \\
     47 &   950.316 & $-1.903$ & 1 &  W3R4 & blend \\
     48 &   950.401 & $-1.564$ & 1 &  W3Q3 & blend \\
     49 &   950.820 & $-2.003$ & 1 &  L14R3 & blend \\
     50 &   951.672 & $-1.896$ & 1 &  W3P3 \\
     51 &   952.256 & $-2.416$ & 1 &  L15P5 & blend \\
     52 &   952.276 & $-3.151$ & 1 &  L14P3 & blend \\
     53 &   952.758 & $-1.565$ & 1 &  W3Q4 & blend \\
     54 &   952.807 & $-1.762$ & 1 &  W3R5 & blend \\
     55 &   954.419 & $-1.860$ & 1 &  L13R0 & blend \\
     56 &   954.475 & $-1.873$ & 1 &  W3P4 & blend \\
     57 &   955.067 & $-2.024$ & 1 &  L13R1 \\
     58 &   955.682 & $-1.567$ & 1 &  W3Q5 \\
     59 &   955.711 & $-2.374$ & 1 &  L13P1 \\
     60 &   956.581 & $-2.064$ & 1 &  L13R2 \\
     61 &   957.654 & $-2.316$ & 1 &  L13P2 \\
     62 &   958.949 & $-2.087$ & 1 &  L13R3 \\
%article only:
\shrinkheight{-2cm}
     63 &   960.452 & $-2.307$ & 1 &  L13P3 \\
     64 &   962.978 & $-1.889$ & 1 &  L12R0 \\
     65 &   963.609 & $-2.163$ & 1 &  L12R1 & blend \\
     66 &   964.312 & $-2.300$ & 1 &  L12P1 \\
     67 &   964.988 & $-2.821$ & 1 &  L12R2 & blend \\
     68 &   964.988 & $-1.162$ & 1 &  W2R0 & blend \\
     69 &   965.067 & $-1.453$ & 1 &  W2R1 & blend \\
     70 &   965.793 & $-1.491$ & 1 &  W2R2 \\
     71 &   966.097 & $-1.457$ & 1 &  W2Q1 \\
     72 &   966.272 & $-2.189$ & 1 &  L12P2 \\
     73 &   966.780 & $-2.104$ & 1 &  W2R3 \\
     74 &   967.278 & $-1.458$ & 1 &  W2Q2 \\
     75 &   967.674 & $-1.667$ & 1 &  L12R3 \\
%referee version only:
%\shrinkheight{-1.5cm}
     76 &   968.293 & $-2.148$ & 1 &  W2P2 \\
     77 &   969.047 & $-1.458$ & 1 &  W2Q3 & blend \\
     78 &   969.086 & $-2.095$ & 1 &  L12P3 & blend \\
     79 &   970.560 & $-2.022$ & 1 &  W2P3 \\
     80 &   971.386 & $-1.460$ & 1 &  W2Q4 \\
     81 &   971.984 & $-1.705$ & 1 &  L11R0 & weak \\
     82 &   973.348 & $-2.231$ & 1 &  L11P1 \\
     83 &   974.156 & $-1.903$ & 1 &  L11R2 \\
     84 &   975.343 & $-2.178$ & 1 &  L11P2 & weak \\
     85 &   978.216 & $-2.175$ & 1 &  L11P3 \\
     86 &   981.441 & $-1.690$ & 1 &  L10R0 \\
     87 &   982.074 & $-1.876$ & 1 &  L10R1 \\
     88 &   982.834 & $-2.169$ & 1 &  L10P1 \\
     89 &   983.595 & $-1.950$ & 1 &  L10R2 \\
     90 &   984.866 & $-2.091$ & 1 &  L10P2 \\
     91 &   985.632 & $-1.157$ & 1,2 &  W1R0 & blend \\
     92 &   985.651 & $-1.474$ & 1,2 &  W1R1 & blend \\
     93 &   985.967 & $-2.108$ & 1 &  L10R3 \\
     94 &   986.246 & $-1.570$ & 1 &  W1R2 \\
     95 &   986.798 & $-1.439$ & 1,2 &  W1Q1 \\
     96 &   987.450 & $-1.578$ & 1 &  W1R3 \\
     97 &   987.770 & $-2.056$ & 1 &  L10P3 \\
     98 &   987.978 & $-1.440$ & 1 &  W1Q2 \\
     99 &   991.388 & $-1.939$ & 1 &  W1P3 & blend \\
    100 &   991.394 & $-1.587$ & 1 &  L9R0 & blend \\
    101 &   992.018 & $-1.441$ & 1 &  W1Q4 & blend \\
    102 &   992.022 & $-1.745$ & 1 &  L9R1 & blend \\
    103 &   992.052 & $-1.638$ & 1 &  W1R5 & blend \\
    104 &   992.813 & $-2.116$ & 1 &  L9P1 \\
    105 &   993.492 & $-1.959$ & 1 &  L10R5 & blend, weak \\
    106 &   993.549 & $-1.780$ & 1 &  L9R2 & blend \\
    107 &   994.229 & $-1.885$ & 1 &  W1P4 \\
    108 &   994.876 & $-2.066$ & 1 &  L9P2 & blend \\
    109 &   994.924 & $-1.442$ & 1 &  W1Q5 & blend \\
    110 &   995.975 & $-1.793$ & 1 &  L9R3 \\
    111 &   997.829 & $-2.063$ & 1,2 &  L9P3 & see Add.\\
    112 &  1001.826 & $-1.575$ & 1 &  L8R0 \\
    113 &  1002.457 & $-1.742$ & 1,2 &  L8R1 \\
    114 &  1003.304 & $-2.076$ & 1 &  L8P1 \\
%referee version only:
%\shrinkheight{-1.5cm}
    115 &  1003.989 & $-1.785$ & 1 &  L8R2 \\
    116 &  1005.397 & $-2.009$ & 1 &  L8P2 \\
    117 &  1006.346 & $-2.080$ & 1 &  L9P5 & blend \\
    118 &  1006.418 & $-1.812$ & 1 &  L8R3 & blend \\
    119 &  1008.392 & $-1.991$ & 1 &  L8P3 & blend \\
    120 &  1008.502 & $-1.672$ & 1 &  W0R1 & blend \\
    121 &  1008.553 & $-1.349$ & 1 &  W0R0 & blend \\
%article only:
\shrinkheight{-2cm}
    122 &  1009.030 & $-1.783$ & 1 &  W0R2 \\
    123 &  1009.721 & $-2.866$ & 1 &  L8R4 & blend \\
    124 &  1009.772 & $-1.623$ & 1,2 &  W0Q1 & blend \\
    125 &  1010.132 & $-1.833$ & 1 &  W0R3 \\
    126 &  1010.941 & $-1.623$ & 1 &  W0Q2 \\
    127 &  1012.173 & $-2.276$ & 1 &  W0P2 & blend \\
    128 &  1012.261 & $-1.983$ & 1 &  L8P4 & blend \\
    129 &  1012.681 & $-1.625$ & 1 &  W0Q3 \\
    130 &  1012.822 & $-1.527$ & 1 &  L7R0 \\
    131 &  1013.434 & $-1.688$ & 1,2 &  L7R1 & blend \\
    132 &  1013.480 & $-1.821$ & 1 &  W0R5 & blend \\
    133 &  1014.334 & $-2.051$ & 1 &  L7P1 \\
    134 &  1014.509 & $-2.107$ & 1 &  W0P3 \\
    135 &  1014.977 & $-1.724$ & 1 &  L7R2 & blend \\
    136 &  1014.980 & $-1.625$ & 1 &  W0Q4 & blend \\
    137 &  1016.472 & $-1.996$ & 1 &  L7P2 \\
    138 &  1017.390 & $-2.036$ & 1 &  W0P4 & blend \\
    139 &  1017.428 & $-1.735$ & 1 &  L7R3 & blend \\
    140 &  1019.506 & $-1.991$ & 1 &  L7P3 \\
    141 &  1023.443 & $-1.996$ & 1 &  L7P4 \\
    142 &  1024.364 & $-1.540$ & 1 &  L6R0 \\
    143 &  1024.986 & $-1.703$ & 1 &  L6R1 & blend \\
    144 &  1024.991 & $-1.842$ & 1 &  L7R5 & blend \\
    145 &  1026.532 & $-1.740$ & 1 &  L6R2 \\
    146 &  1028.103 & $-1.983$ & 1 &  L6P2 \\
    147 &  1028.986 & $-1.752$ & 1 &  L6R3 \\
    148 &  1031.191 & $-1.966$ & 1 &  L6P3 \\
    149 &  1032.356 & $-1.757$ & 1 &  L6R4 \\
    150 &  1036.546 & $-1.567$ & 1 &  L5R0 & blend \\
    151 &  1036.546 & $-1.836$ & 1 &  L6R5 & blend \\
    152 &  1037.146 & $-1.733$ & 1 &  L5R1 & blend \\
%referee version only:
%\shrinkheight{-1.5cm}
    153 &  1038.156 & $-2.068$ & 1 &  L5P1 \\
    154 &  1038.690 & $-1.770$ & 1 &  L5R2 \\
    155 &  1040.367 & $-2.001$ & 1 &  L5P2 \\
    156 &  1041.156 & $-1.785$ & 1 &  L5R3 \\
    157 &  1043.498 & $-1.983$ & 1 &  L5P3 \\
    158 &  1044.546 & $-1.791$ & 1 &  L5R4 & weak \\
    159 &  1047.554 & $-1.979$ & 1 &  L5P4 \\
    160 &  1049.366 & $-1.629$ & 1 &  L4R0 \\
    161 &  1049.958 & $-1.796$ & 1 &  L4R1 \\
    162 &  1051.031 & $-2.125$ & 1 &  L4P1 \\
    163 &  1051.497 & $-1.833$ & 1 &  L4R2 \\
    164 &  1053.281 & $-2.056$ & 1 &  L4P2 \\
    165 &  1053.976 & $-1.848$ & 1 &  L4R3 \\
    166 &  1056.469 & $-2.037$ & 1 &  L4P3 \\
    167 &  1057.379 & $-1.854$ & 1 &  L4R4 & weak \\
    168 &  1060.580 & $-2.032$ & 1 &  L4P4 & weak \\
    169 &  1062.883 & $-1.740$ & 1 &  L3R0 & blend \\
    170 &  1063.460 & $-1.907$ & 1 &  L3R1 \\
    171 &  1064.606 & $-2.234$ & 1 &  L3P1 \\
    172 &  1064.995 & $-1.947$ & 1 &  L3R2 \\
    173 &  1066.900 & $-2.163$ & 1 &  L3P2 \\
    174 &  1067.478 & $-1.963$ & 1 &  L3R3 \\
    175 &  1070.142 & $-2.142$ & 1 &  L3P3 \\
    176 &  1077.138 & $-1.924$ & 1 &  L2R0 \\
    177 &  1077.698 & $-2.092$ & 1 &  L2R1 \\
    178 &  1078.925 & $-2.415$ & 1 &  L2P1 \\
    179 &  1079.226 & $-2.132$ & 1 &  L2R2 \\
    180 &  1081.265 & $-2.344$ & 1 &  L2P2 \\
    181 &  1081.710 & $-2.147$ & 1 &  L2R3 & blend \\
    182 &  1084.559 & $-2.321$ & 1 &  L2P3 \\
    183 &  1092.194 & $-2.225$ & 1 &  L1R0 \\
    184 &  1092.732 & $-2.395$ & 1 &  L1R1 \\
    185 &  1093.955 & $-2.310$ & 1 &  L2P5 & weak \\
    186 &  1094.052 & $-2.714$ & 1 &  L1P1 \\
    187 &  1094.244 & $-2.436$ & 1 &  L1R2 \\
    188 &  1096.439 & $-2.642$ & 1 &  L1P2 \\
    189 &  1096.725 & $-2.451$ & 1 &  L1R3 & blend \\
    190 &  1099.788 & $-2.620$ & 1 &  L1P3 \\
    191 &  1100.165 & $-2.460$ & 1 &  L1R4 & weak \\
    192 &  1108.128 & $-2.762$ & 1 &  L0R0 \\
    193 &  1108.634 & $-2.932$ & 1 &  L0R1 \\
    194 &  1110.063 & $-3.250$ & 1 &  L0P1 & blend \\
    195 &  1110.120 & $-2.971$ & 1 &  L0R2 & blend \\
    196 &  1112.495 & $-3.177$ & 1 &  L0P2 & blend \\
    197 &  1112.584 & $-2.991$ & 1 &  L0R3 & blend \\
    198 &  1115.896 & $-3.153$ & 1 &  L0P3 \\
\hline
\end{supertabular}
} % \small

\subsection{Interstellar metal lines}

Most of the metal lines can be observed in both interstellar
components with the second component being only slightly weaker than
the main component. High resolution spectra do show more components
(Spitzer \& Fitzpatrick \cite{spitzer2}), but in the {\it ORFEUS}
echelle spectra only two well separated components are seen. The
separation is best seen in the Ar\,I lines $\lambda$1048 and
$\lambda$1067 and the N\,I triplett $\lambda$1134. Interstellar O\,VI
at $\lambda$1032 and $\lambda$1037 appears quite broad. Widmann
estimated a $N(O\,VI)\,=\,(0.99\pm0.15$)\,10$^{14}$\,cm$^{-2}$ from these
{\it ORFEUS} echelle spectra (Widmann et al. \cite{widmann1}; Widmann
\cite{widmann2}).

\paragraph{}
{\small

\tablecaption{Table of identified interstellar metal lines. For
unresolved doublets and triplets an average wavelength and a calculated
effective $\log(gf)$ are given. VC is the velocity component as given in
Table\,1.}

\tablehead{
  \hline
  No. & $\lambda$ [{\AA}] & $\log(gf)$ & VC & Elem. & Remarks \\
  \hline
}

\tabletail{
  \hline
  \multicolumn{5}{l}{Table 3, continued ...} \\
  \hline
}

\tablelasttail{
  \hline
}
\begin{supertabular}{rrrlll}
    199 &   919.658 & $-2.391$ & 1,2 & O I & triplet, blend \\
%article only:
\shrinkheight{-0.5cm}
%referee version only:
%\shrinkheight{-0.5cm}
    200 &   924.952 & $-2.099$ & 1,2 & O I & triplet \\
    201 &   925.442 & $-2.752$ & 1,2 & O I & comp. 2 ? \\
    202 &   926.212 & $-0.359$ & 1,2 & Fe II & blend \\
    203 &   929.517 & $-1.928$ & 1,2 & O I & triplet, blend \\
    204 &   930.257 & $-2.571$ & 1,2 & O I & weak \\
    205 &   936.629 & $-1.729$ & 1,2 & O I & triplet \\
    206 &   945.191 & $-0.565$ & 1,2 & C I & weak \\
    207 &   946.978 & $-0.230$ & 1,2 & S II & blend \\
    208 &   948.686 & $-1.492$ & 1,2 & O I & triplet, blend \\
    209 &   950.112 & $-1.714$ & 1 & O I* & doublet, blend \\
    210 &   950.885 & $-2.105$ & 1,2 & O I & blend \\
    211 &   952.303 & $-2.152$ & 1,2 & N I & blend \\
    212 &   952.415 & $-2.206$ & 1,2 & N I & blend \\
    213 &   953.415 & $-1.371$ & 1,2 & N I & blend \\
    214 &   953.655 & $-1.090$ & 1,2 & N I & blend \\
%referee version only:
%\shrinkheight{-1.5cm}
    215 &   953.970 & $-0.983$ & 1,2 & N I & blend \\
    216 &   954.104 & $-1.752$ & 1,2 & N I & blend \\
    217 &   961.041 & $-0.457$ & 1,2 & P II & weak \\
    218 &   963.801 & $ 0.164$ & 1,2 & P II & blend \\
    219 &   963.990 & $-1.134$ & 1,2 & N I & comp.2 blend \\
    220 &   964.626 & $-1.326$ & 1,2 & N I &  \\
    221 &   965.041 & $-1.634$ & 1,2 & N I & blend \\
    222 &   971.738 & $-1.131$ & 1,2 & O I & triplet \\
    223 &   977.020 & $-0.118$ & 1 & C III & blend \\
    224 &  1020.699 & $-1.248$ & 1,2 & Si II &  \\
    225 &  1031.926 & $-0.575$ & 1,2 & O VI &  \\
    226 &  1036.337 & $-0.609$ & 1,2 & C II &  \\
    227 &  1037.018 & $-0.308$ & 1,2 & C II* & blend \\
    228 &  1037.617 & $-0.879$ & 1,2 & O VI & blend \\
    229 &  1039.230 & $-1.337$ & 1,2 & O I &  \\
    230 &  1048.220 & $-0.612$ & 1,2 & Ar I &  \\
    231 &  1055.262 & $-1.097$ & 1,2 & Fe II & weak \\
    232 &  1063.176 & $-0.222$ & 1,2 & Fe II & blend \\
    233 &  1063.972 & $-1.347$ & 1,2 & Fe II & weak \\
    234 &  1066.660 & $-1.177$ & 1,2 & Ar I &  \\
    235 &  1081.875 & $-0.854$ & 1,2 & Fe II & blend \\
    236 &  1083.990 & $-0.987$ & 1,2 & N II &  \\
    237 &  1096.877 & $-0.495$ & 1,2 & Fe II & blend \\
    238 &  1112.048 & $-1.040$ & 1,2 & Fe II & weak, blend \\
    239 &  1121.975 & $-0.699$ & 1,2 & Fe II &  \\
    240 &  1122.526 & $-0.149$ & 1,2 & Fe III &  \\
    241 &  1125.448 & $-0.959$ & 1,2 & Fe II &  \\
    242 &  1133.665 & $-1.222$ & 1,2 & Fe II &  \\
    243 &  1134.165 & $-1.270$ & 1,2 & N I & comp.1 blend \\
    244 &  1134.415 & $-0.969$ & 1,2 & N I & comp.2 blend \\
    245 &  1134.980 & $-0.793$ & 1,2 & N I &  \\
    246 &  1143.226 & $-0.876$ & 1,2 & Fe II &  \\
    247 &  1144.938 & $ 0.021$ & 1,2 & Fe II &  \\
    248 &  1152.818 & $-0.627$ & 1,2 & P II &  \\
    249 &  1190.416 & $-0.301$ & 1,2 & Si II &  \\
    250 &  1193.290 & $-0.001$ & 1,2 & Si II &  \\
    251 &  1199.550 & $-0.275$ & 1,2 & N I &  \\
    252 &  1200.223 & $-0.451$ & 1,2 & N I &  \\
    253 &  1200.710 & $-0.752$ & 1,2 & N I &  \\
    254 &  1206.500 & $ 0.223$ & 1,2 & Si III & blend \\
    255 &  1250.584 & $-1.661$ & 1,2 & S II &  \\
    256 &  1253.811 & $-1.361$ & 1,2 & S II &  \\
    257 &  1259.519 & $-1.187$ & 1,2 & S II &  \\
    258 &  1260.422 & $ 0.304$ & 1,2 & Si II & blend \\
    259 &  1260.533 & $-0.602$ & 1,2 & Fe II & blend \\
    260 &  1277.245 & $-1.015$ & 1 & C I & weak \\
    261 &  1302.168 & $-0.612$ & 1,2 & O I &  \\
    262 &  1304.370 & $-0.531$ & 1,2 & Si II &  \\
    263 &  1334.532 & $-0.593$ & 1,2 & C II &  \\
%article only:
\shrinkheight{-1cm}
    264 &  1335.690 & $-0.292$ & 1,2 & C II* & doublet \\
\hline
\end{supertabular}
} % \small

\subsection{Lyman series lines}

Of the Lyman series 14 lines are detectable, from which the lines below
the $\lambda$915.824 line are not separated, so that this line marks
the interstellar Lyman limit towards HD\,93521.

\paragraph{}
{\small

\tablecaption{Table of the identified interstellar Lyman-series (H\,I).
All lines are doublets with a maximum separation of 5.4\,m{\AA} (at
$\lambda$1216). The resulting wavelength and $\log(gf)$ is given. VC is
the velocity component as given in Table\,1.}

\tablehead{
  \hline
  No. & $\lambda$ [{\AA}] & $\log(gf)$ & VC &  Lyman & Remarks \\
      &                   &            &    &  Name  &   \\
  \hline
}

\tabletail{
  \hline
  \multicolumn{5}{l}{Table 4, continued ...} \\
  \hline
}

\tablelasttail{
  \hline
}
\begin{supertabular}{rrrlrl}
    265 &   915.824 & $-3.029$ & 1 & 14 &  \\
    266 &   916.429 & $-2.938$ & 1 & 13 &  \\
    267 &   917.181 & $-2.840$ & 1 & 12 &  \\
    268 &   918.129 & $-2.735$ & 1 & 11 &  \\
    269 &   919.351 & $-2.620$ & 1 & 10 &  \\
    270 &   920.963 & $-2.493$ & 1 & 9 &  \\
    271 &   923.150 & $-2.353$ & 1 & 8 &  \\
    272 &   926.226 & $-2.196$ & 1 & 7 &  \\
    273 &   930.748 & $-2.016$ & 1 & 6 &  \\
    274 &   937.803 & $-1.807$ & 1 & $\epsilon$ &  \\
    275 &   949.743 & $-1.555$ & 1 & $\delta$ &  \\
    276 &   972.537 & $-1.237$ & 1 & $\gamma$ &  \\
    277 &  1025.722 & $-0.801$ & 1 & $\beta$  &  \\
    278 &  1215.670 & $-0.079$ & 1 & $\alpha$ &  \\
    279 &  1215.670 & $-0.079$ & 4 & $\alpha$ & geocoronal \\
\hline
\end{supertabular}
} % \small

\subsection{Stellar lines}

The two N\,V-lines at $\lambda$$\lambda$1239/1243 have a pronounced
P-Cygni profile. Within the absorption part both lines show two
significant narrow absorption components at $-270$ and
$-340$\,km\,s$^{-1}$. These narrow absorption components could be an
indication for a disk in the wind of HD\,93521 (Bjorkman et al.
\cite{bjorkman}). The same components are also visible in the Si\,IV
doublet at $\lambda$$\lambda$1394/1403, in Si\,IV $\lambda$1073 and in
Si\,III $\lambda$1206.

The strong Si\,III $\lambda$$\lambda$1300 triplets are seen as wind
absorption lines (Massa \cite{massa}). They appear at a range between
$-60$\,km\,s$^{-1}$ and $-100$\,km\,s$^{-1}$, whereas Massa reports a
value of $-30$\,km\,s$^{-1}$. This difference could be due to some long
term wind variability.

The comparatively strong stellar absorption at 1085\,{\AA} could not be
clearly identified, it is possibly a superposition of different lines.
A candidate is the He\,II $\lambda$1084.9 line, but as the next
lower unblended He\,II line at $\lambda$958.7 is rather weak, the
identification is not sure. There is a Fe\,II resonance line at
$\lambda$1085.0 with a low $\log(gf)=-2.106$, but also non-resonance
lines of Fe\,II and Fe\,III are present in this region. So this line
probably requires a more detailed analysis.

\paragraph{}
{\small

\tablecaption{Table of identified stellar (*) and stellar wind (w)
lines. For the N\,III doublets only the stronger of the two lines is
listed. VC is the velocity component as given in Table\,1.}

\tablehead{
  \hline
  No. & $\lambda$ [{\AA}] & $\log(gf)$ & VC & Elem. & Remarks \\
  \hline
}

\tabletail{
  \hline
  \multicolumn{5}{l}{Table 5, continued ...} \\
  \hline
}

\tablelasttail{
  \hline
}
\begin{supertabular}{rrrlll}
    280 &   977.020 & $-0.118$ & 3 & C III (*) &  \\
    281 &   979.832 & $-0.248$ & 3 & N III* (*) & doublet \\
    282 &   979.905 & $-0.055$ & 3 & N III* (*) & doublet \\
    283 &   989.799 & $-0.671$ & 3 & N III (*) &  \\
    284 &   989.873 & $-0.575$ & 3 & Si II (*) &  \\
    285 &  1031.926 & $-0.575$ & 3 & O VI (*) & P Cygni \\
    286 &  1037.617 & $-0.879$ & 3 & O VI (*) & P Cygni \\
    287 &  1062.662 & $-1.097$ & 3 & S IV (*) &  \\
    288 &  1072.974 & $-0.846$ & 3,5,6 & S IV* (*) &  \\
    289 &  1073.516 & $-1.800$ & 3 & S IV* (*) &  \\
    290 &  1117.977 & $-0.024$ & 3 & P V (*) &  \\
    291 &  1128.008 & $-0.329$ & 3 & P V (*) &  \\
    292 &  1174.933 & $-0.459$ & 3 & C III* (*) &  \\
    293 &  1175.263 & $-0.549$ & 3 & C III* (*) &  \\
    294 &  1175.590 & $-0.679$ & 3 & C III* (*) &  \\
    295 &  1175.711 & $ 0.021$ & 3 & C III* (*) &  \\
    296 &  1175.987 & $-0.549$ & 3 & C III* (*) &  \\
    297 &  1176.370 & $-0.459$ & 3 & C III* (*) &  \\
    298 &  1183.032 & $-0.591$ & 3 & N III* (*) & doublet \\
    299 &  1184.514 & $-0.195$ & 3 & N III* (*) & doublet \\
    300 &  1206.500 & $ 0.223$ & 3,5,6 & Si III (*) &  \\
    301 &  1238.821 & $-0.503$ & 3,5,6 & N V (*) &  \\
    302 &  1242.804 & $-0.806$ & 3,5,6 & N V (*) &  \\
    303 &  1247.383 & $-0.157$ & 3 & C III (*) &  \\
    304 &  1294.545 & $-0.037$ & 7 & Si III* (w) &  \\
    305 &  1296.726 & $-0.127$ & 7 & Si III* (w) &  \\
    306 &  1298.892 & $-0.257$ & 7 & Si III* (w) &  \\
    307 &  1298.946 & $ 0.443$ & 7 & Si III* (w) &  \\
    308 &  1301.149 & $-0.127$ & 7 & Si III* (w) &  \\
    309 &  1303.323 & $-0.037$ & 7 & Si III* (w) &  \\
    310 &  1393.755 & $ 0.012$ & 3,5,6 & Si IV (*) &  \\
    311 &  1402.770 & $-0.292$ & 3,5,6 & Si IV (*) &  \\
\hline
\end{supertabular}
} % \small

\section{Conclusions}

The presented {\it ORFEUS\,II} Echelle spectrum of HD\,93521 shows an
extraordinary rich variety of very sharp interstellar absorption lines,
especially in the wavelength region between 900\,{\AA} and 1200\,{\AA},
which was not very well studied before the two {\it ORFEUS} missions
and for which only now new observation possibilities exist.
Particularly the nearly complete presence of very sharp (FWHM $\approx$
100\,m\AA) H$_{2}$ absorption lines in this spectrum -- which will be
analysed in detail in a forthcoming paper -- makes it well suited as a
reference spectrum for interstellar molecular hydrogen. Molecular
hydrogen is partially visible in the interstellar high velocity
component too. Additionally some stellar lines show narrow absorption
components wich are varying in time and which could be an indication
for a disk.

\begin{acknowledgements}

ORFEUS could only be realized with the support of all our German and
American colleagues and collaborators. The authors wish to thank
Philipp Richter and Klaas S. de Boer for helpful discussions. The
ORFEUS program was supported by DARA grants WE3\,OS\,8501 and
WE2\,QV\,9304 and NASA grant NAG5-696.

\end{acknowledgements}

%\appendix

\section*{Appendix: plots of the spectrum of HD\,93521}

The following plots show the complete ORFEUS\,II Echelle spectrum of
HD\,93521. The numbers shown in squared brackets correspond to the
running numbers given in the Tables\,2\,-\,5. Stellar lines are marked
as (*), stellar wind lines as (w). Non-resonance lines are additionally
marked with an asterisk, *.

The plots show one Echelle order per plot for wavelengths above
1130\,{\AA} (Echelle orders 40 to 49), and half an Echelle order per
plot for Echelle orders 50 to 61 ($\lambda$\,$<$\,1130\,{\AA}) which
includes all H$_{2}$-lines. Thus the wavelength scale changes from
order to order, but the relative wavelength scale (radial velocity
scale) is nearly constant within these two wavelength ranges.

\section*{Addendum}

H$_{2}$-line $\lambda$997.829 (no.\,111) was erroneously identified
with two velocity components. The more probable identification for the
weaker component however is the H$_{2}$-line $\lambda$997.640 W1P5 with
$\log(f)=-1.921$.

\begin{figure*}
\resizebox{\hsize}{7.8cm}{\includegraphics{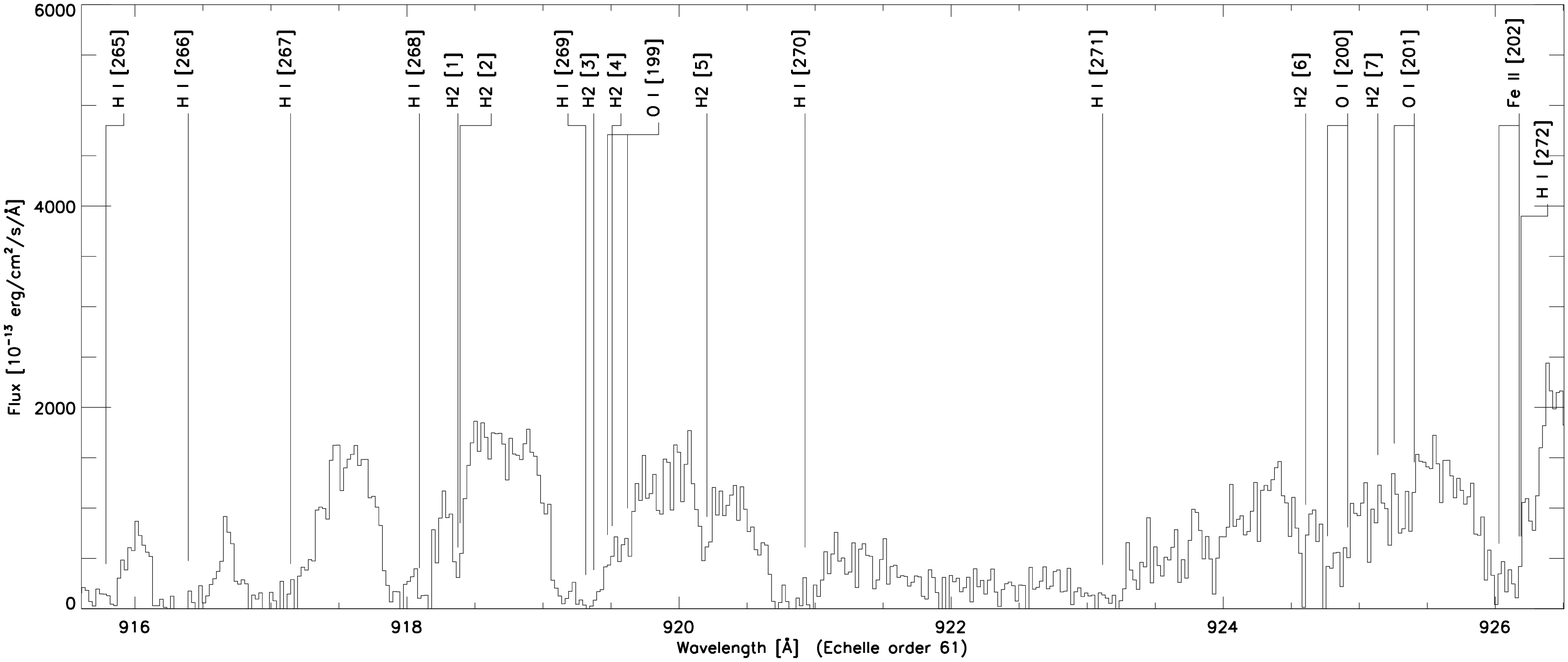}}
\resizebox{\hsize}{7.8cm}{\includegraphics{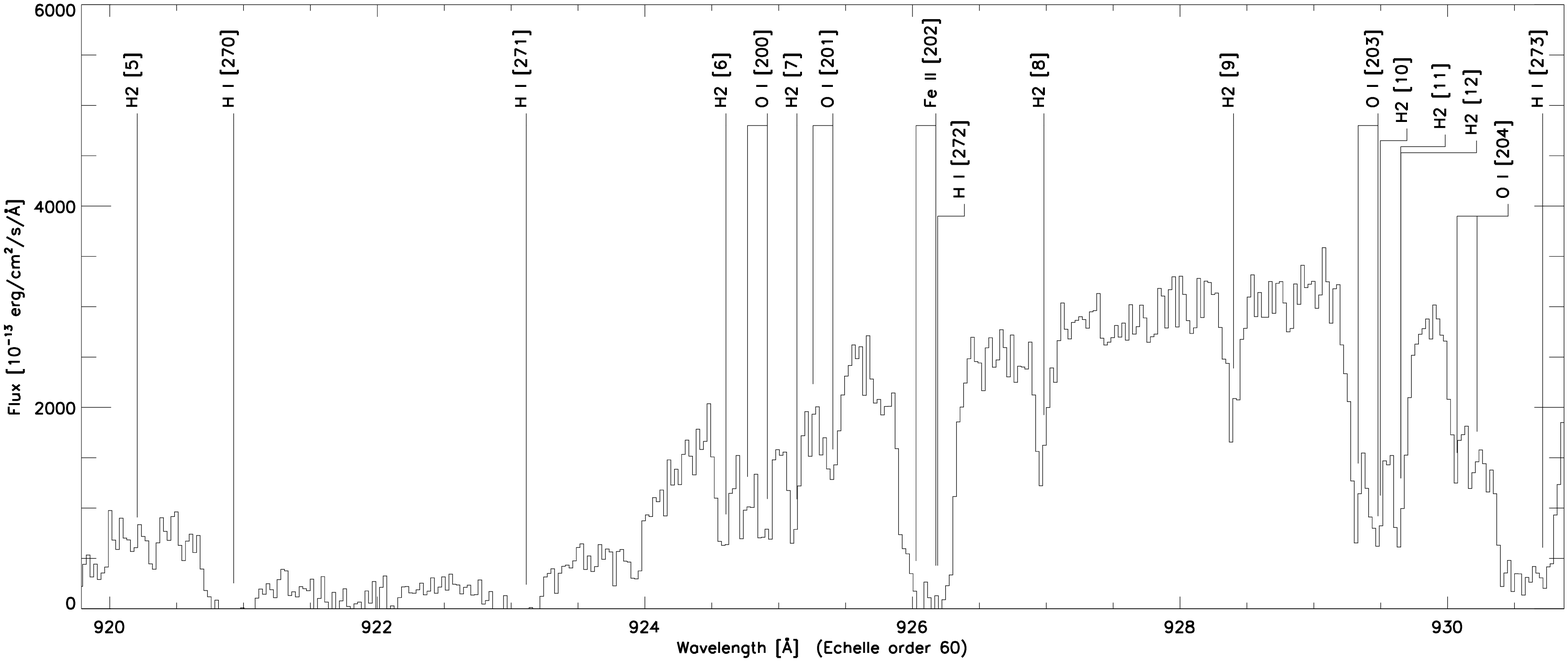}}
\resizebox{\hsize}{7.8cm}{\includegraphics{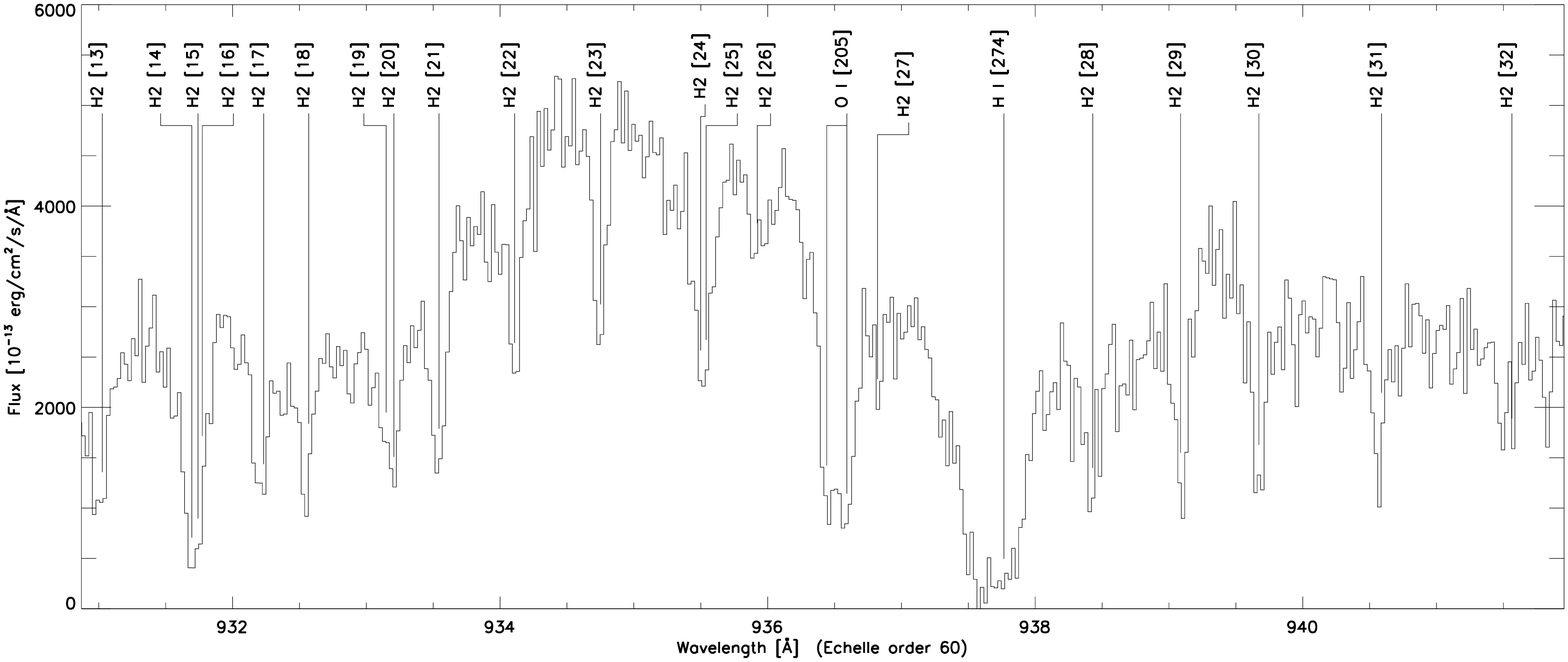}}
\end{figure*}
\begin{figure*}
\resizebox{\hsize}{7.8cm}{\includegraphics{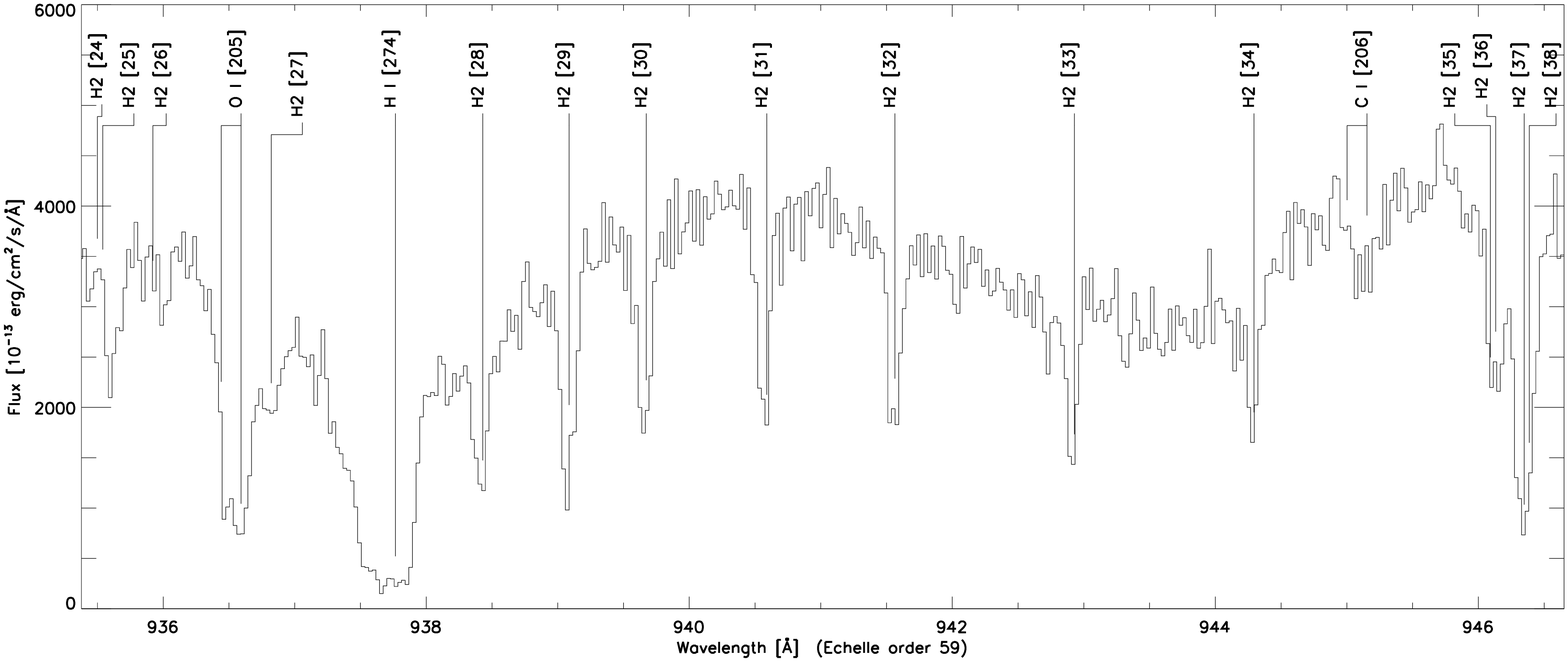}}
\resizebox{\hsize}{7.8cm}{\includegraphics{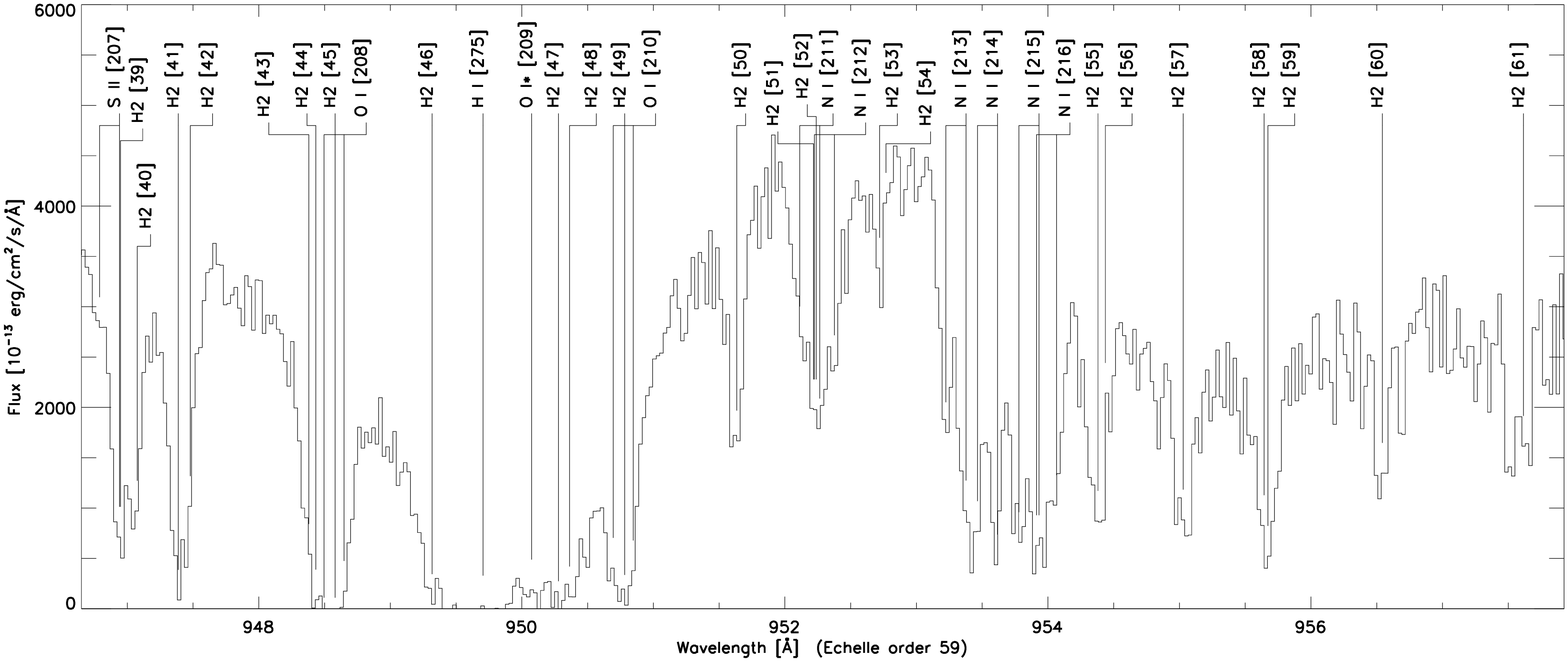}}
\resizebox{\hsize}{7.8cm}{\includegraphics{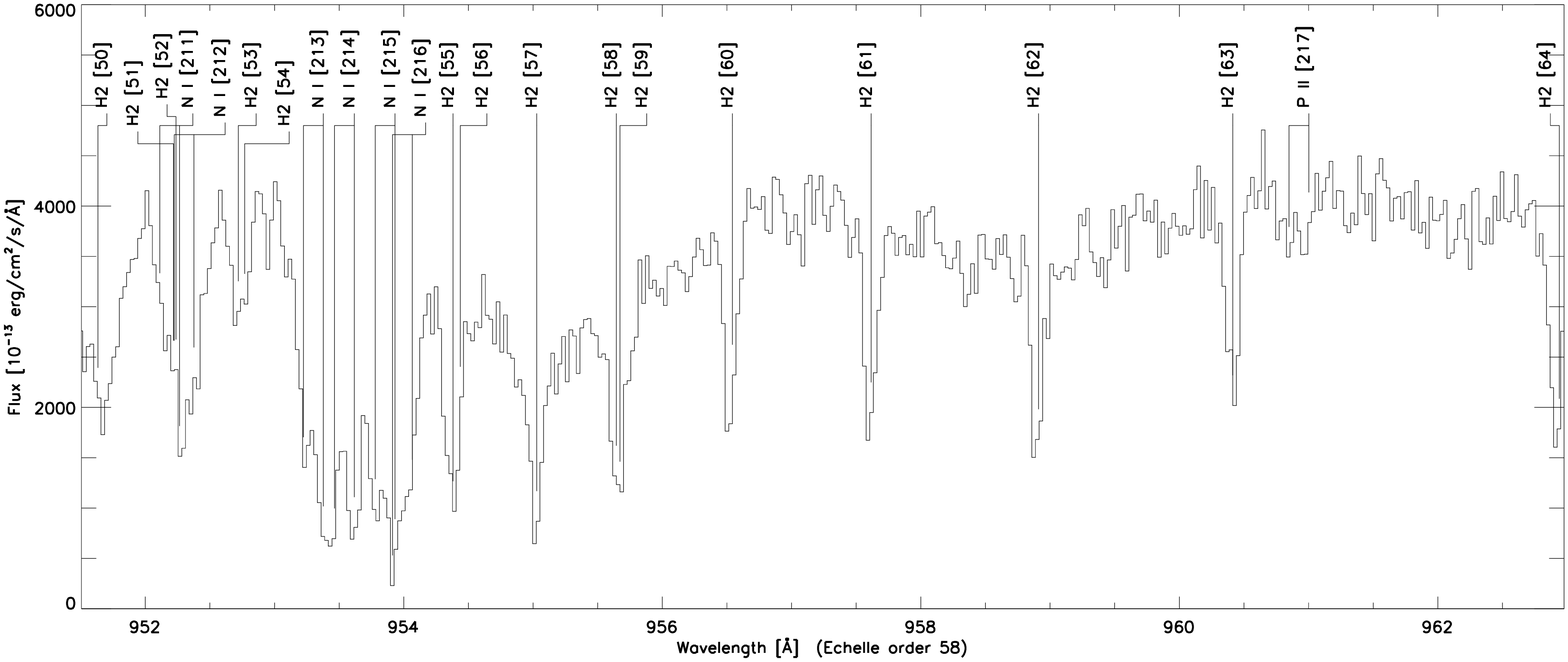}}
\end{figure*}
\begin{figure*}
\resizebox{\hsize}{7.8cm}{\includegraphics{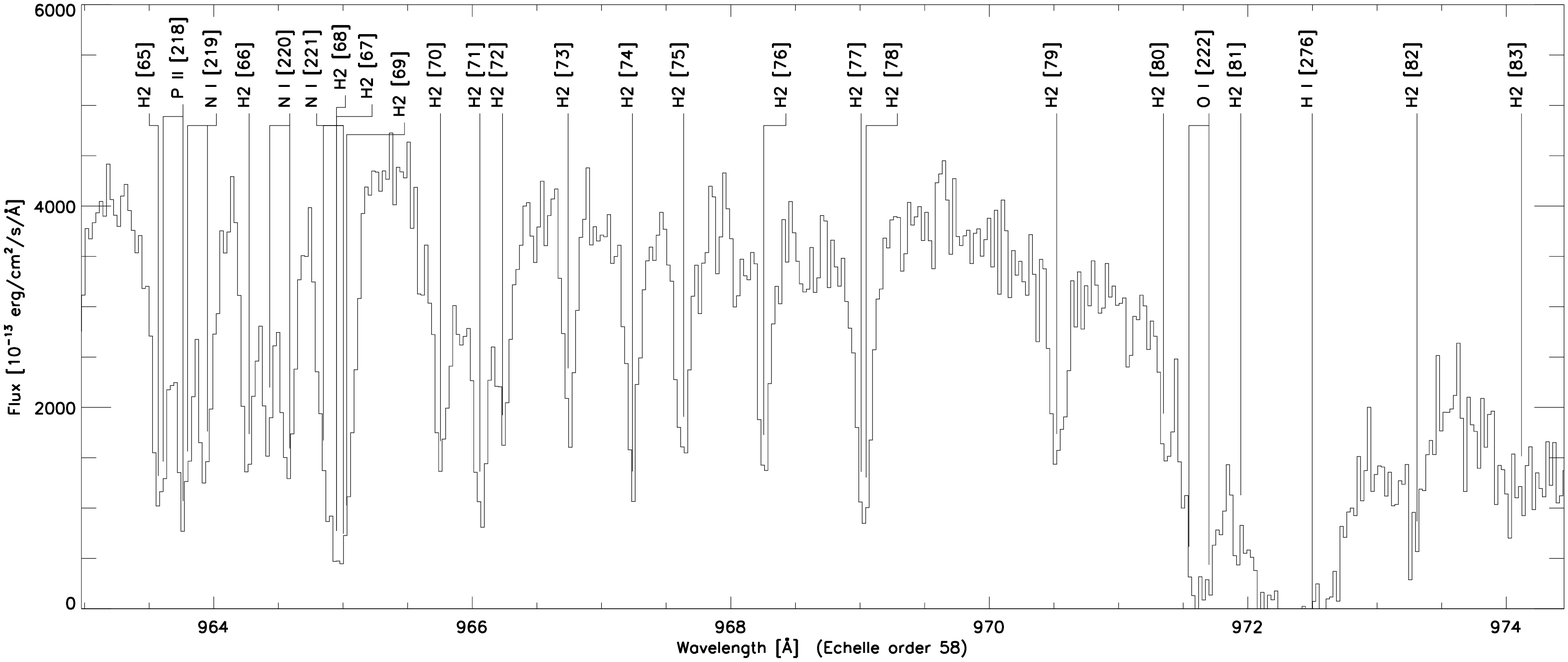}}
\resizebox{\hsize}{7.8cm}{\includegraphics{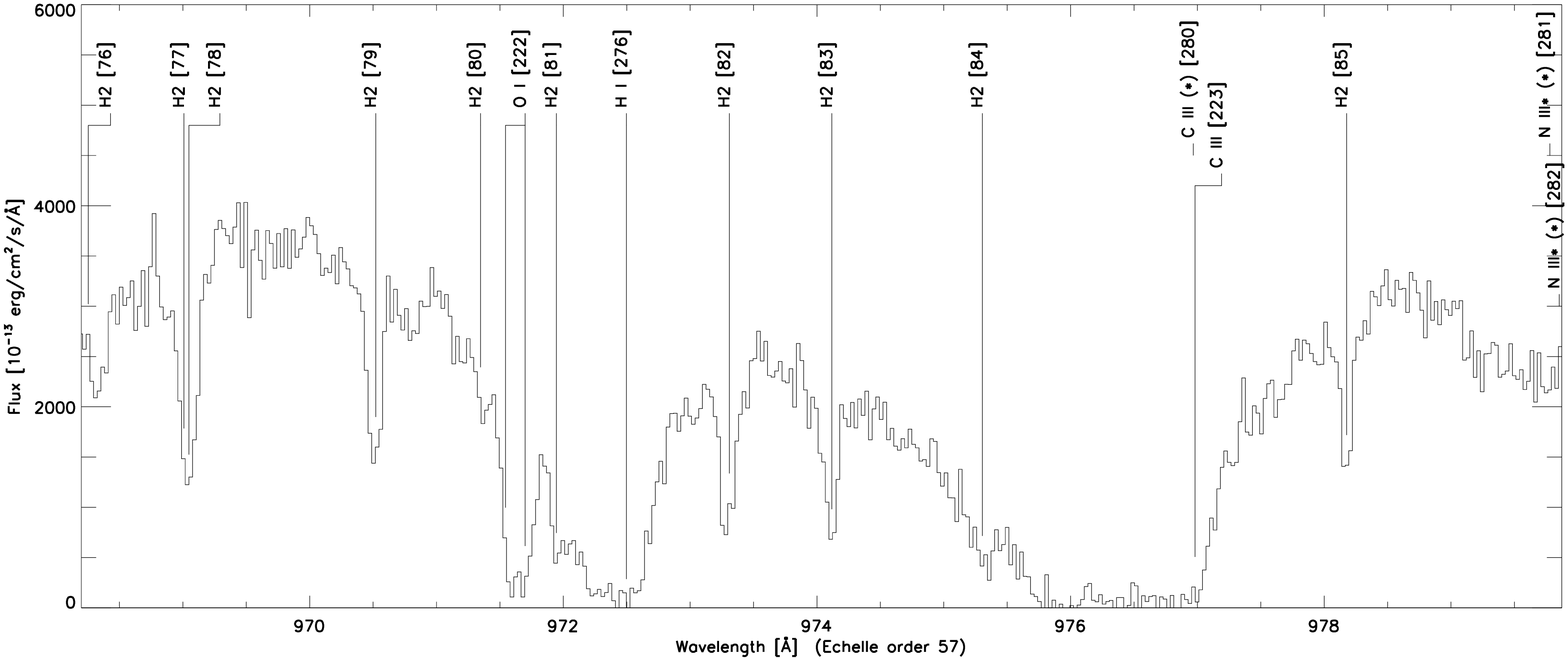}}
\resizebox{\hsize}{7.8cm}{\includegraphics{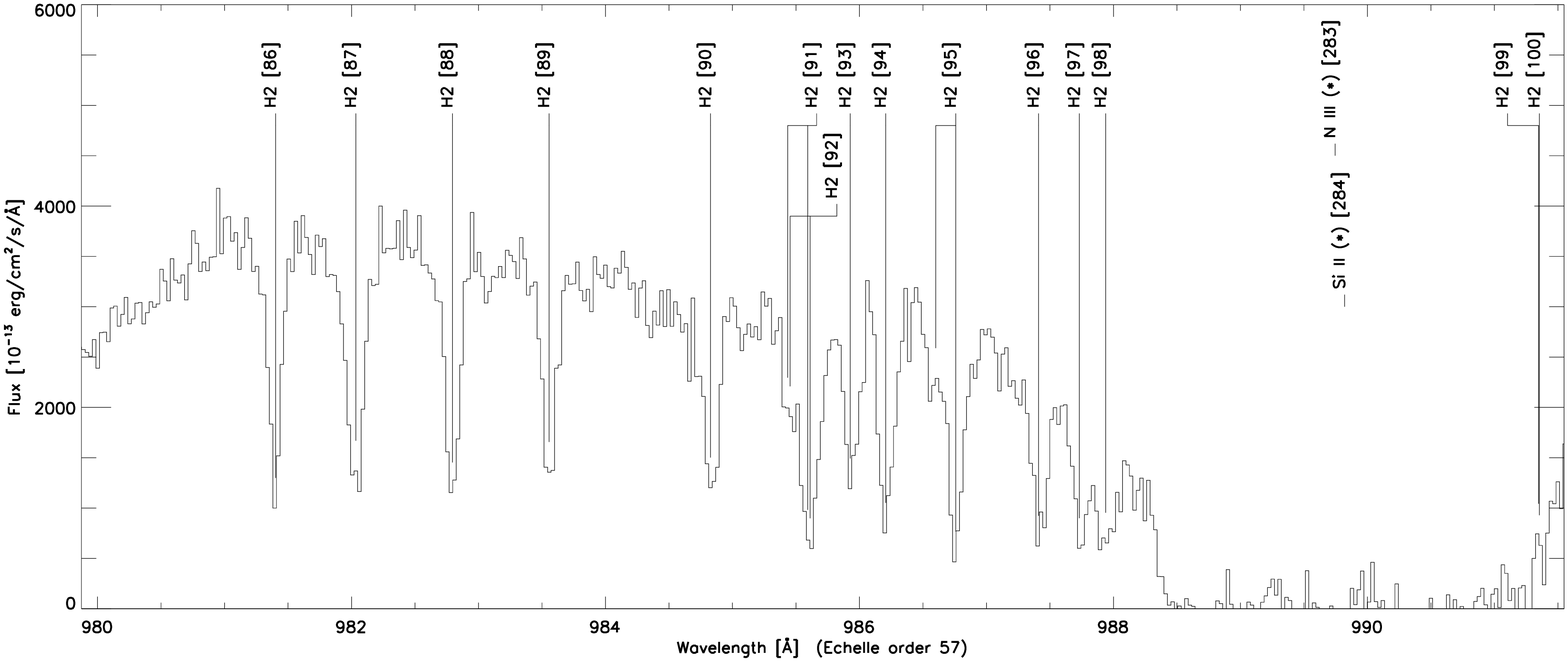}}
\end{figure*}
\begin{figure*}
\resizebox{\hsize}{7.8cm}{\includegraphics{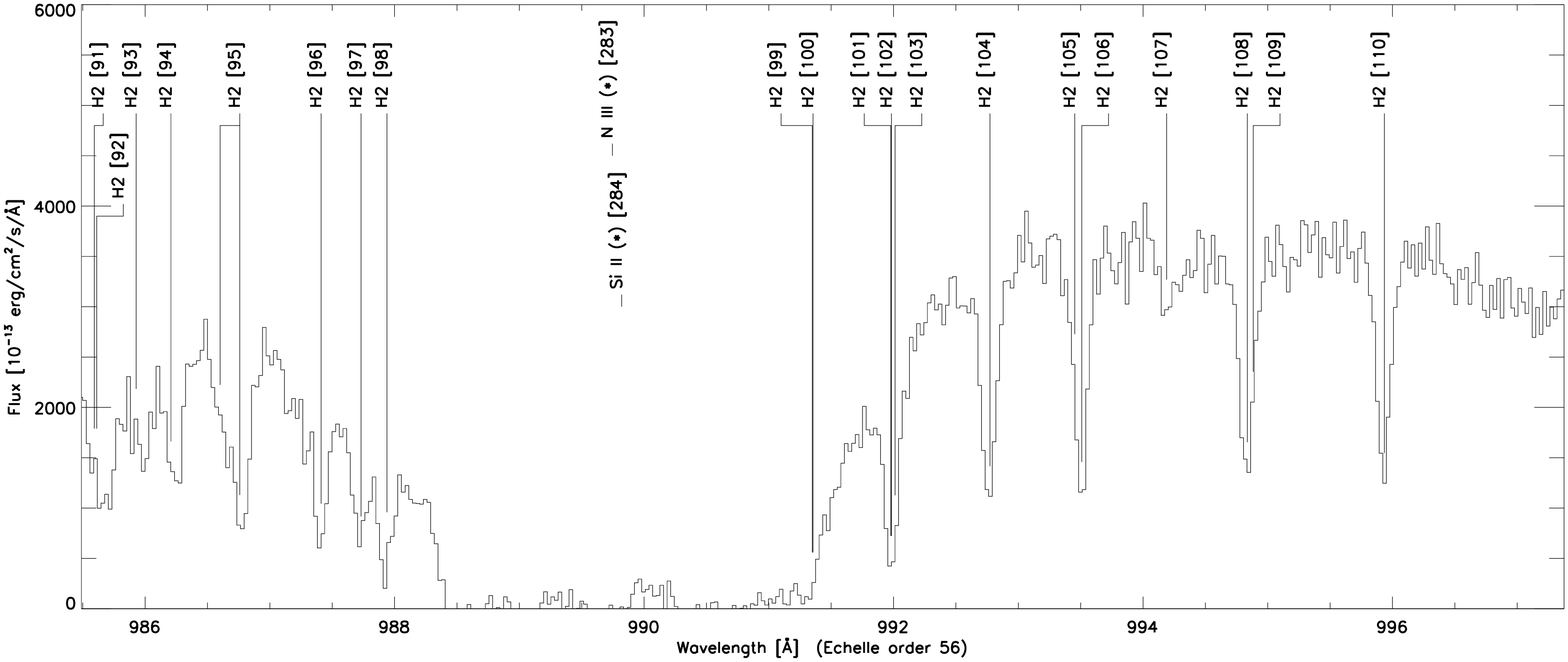}}
\resizebox{\hsize}{7.8cm}{\includegraphics{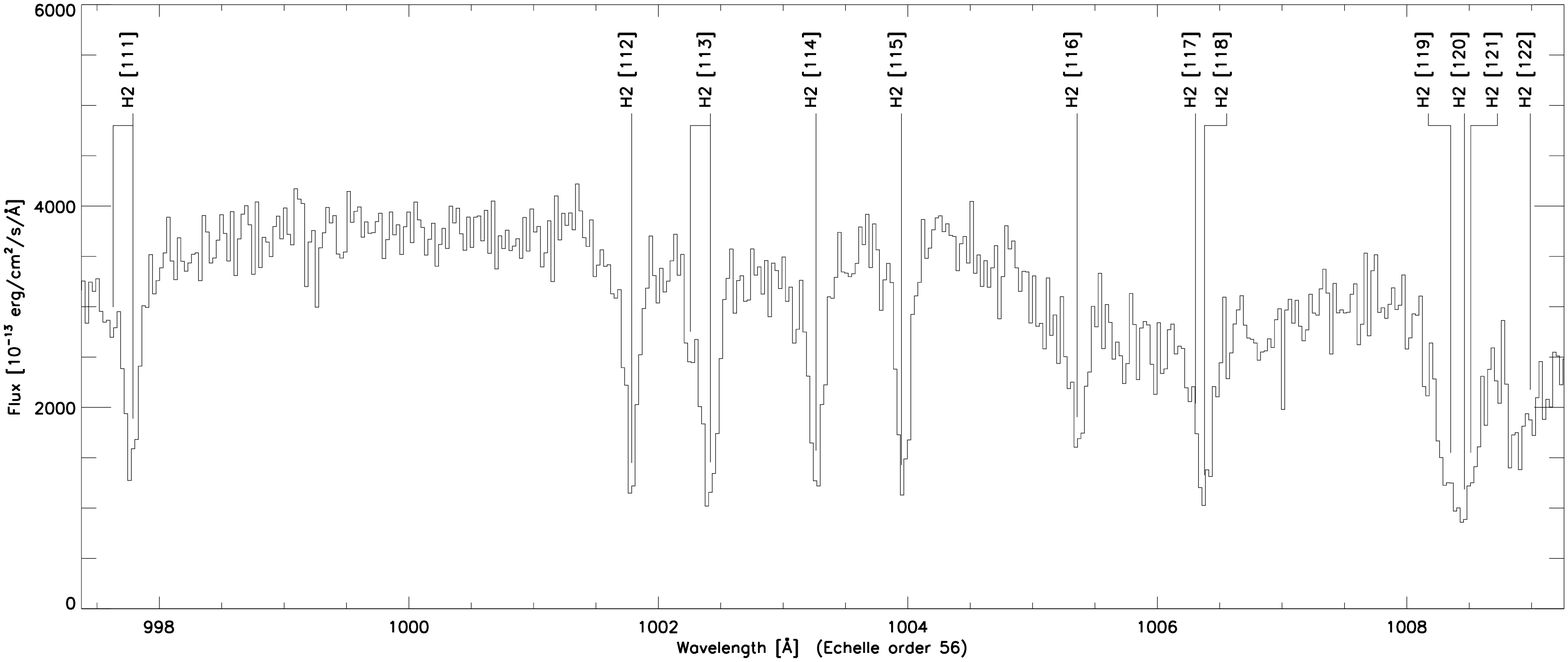}}
\resizebox{\hsize}{7.8cm}{\includegraphics{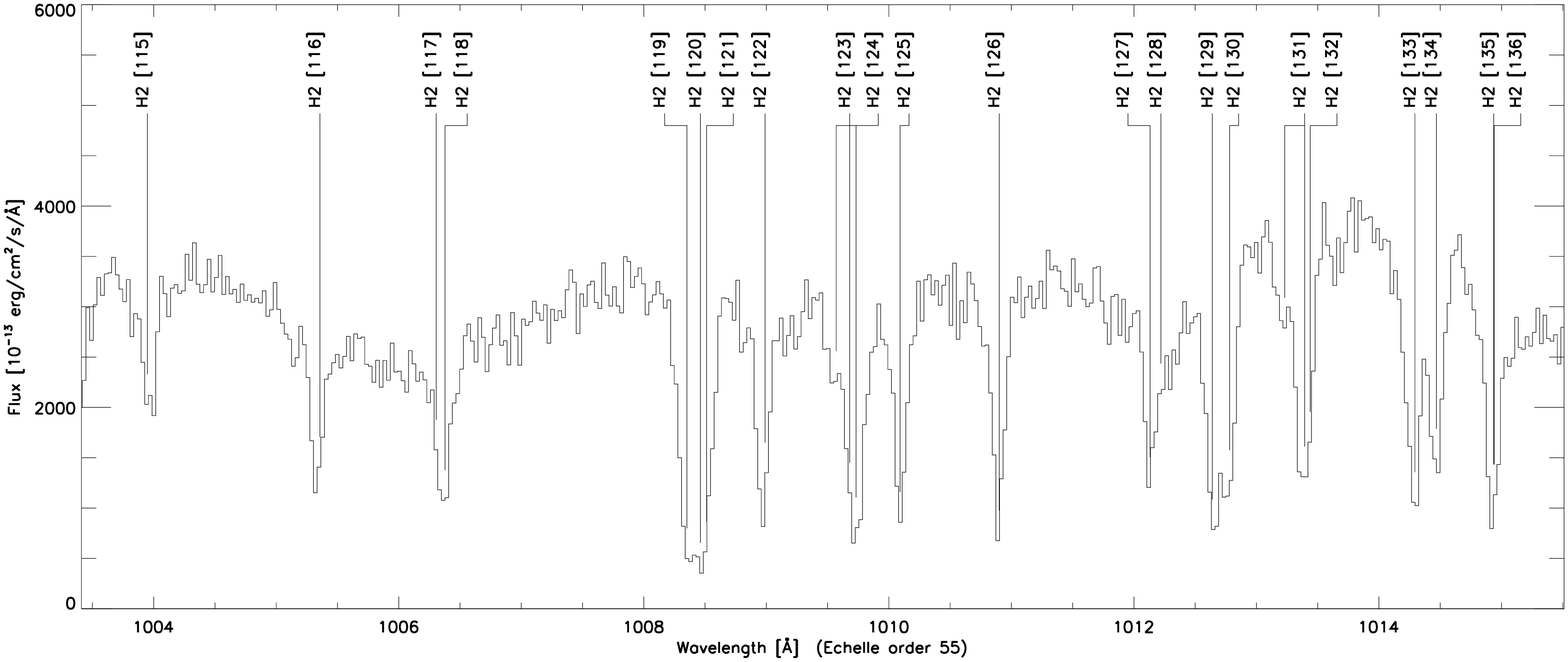}}
\end{figure*}
\begin{figure*}
\resizebox{\hsize}{7.8cm}{\includegraphics{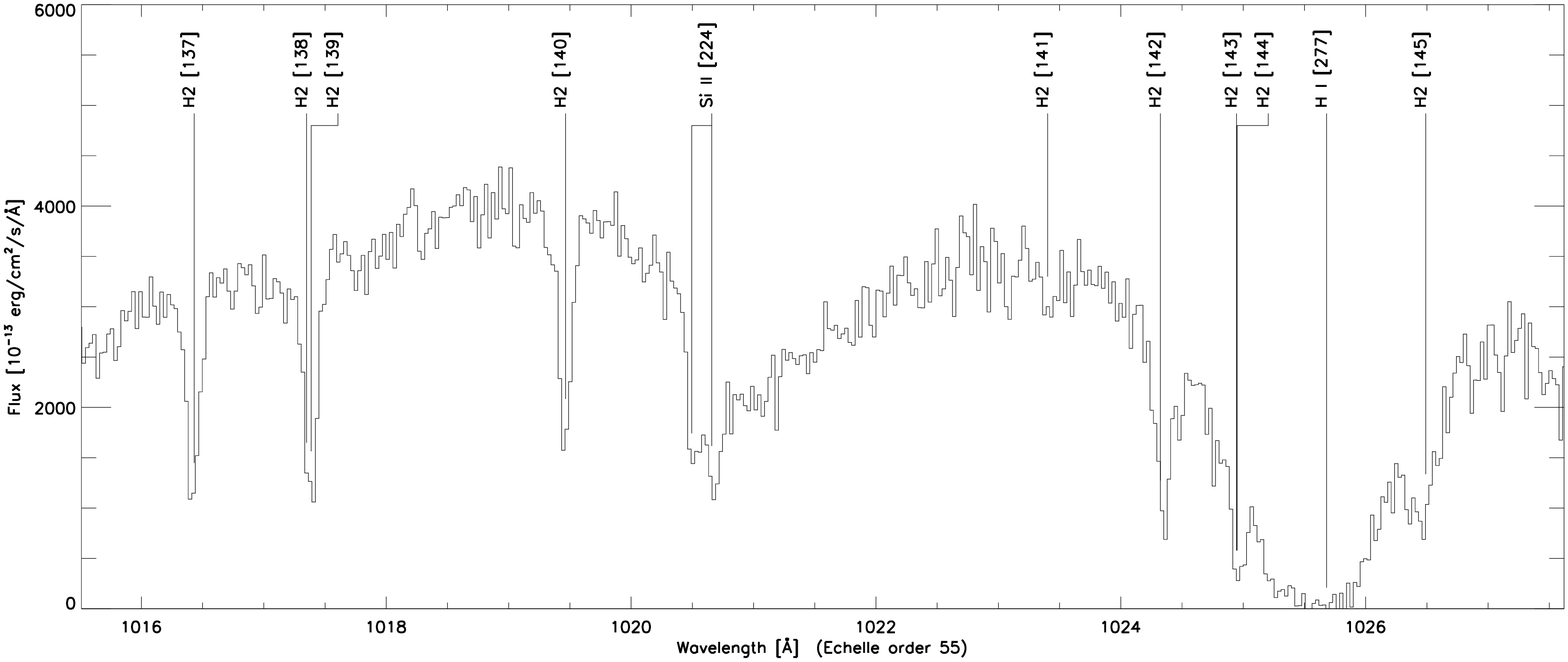}}
\resizebox{\hsize}{7.8cm}{\includegraphics{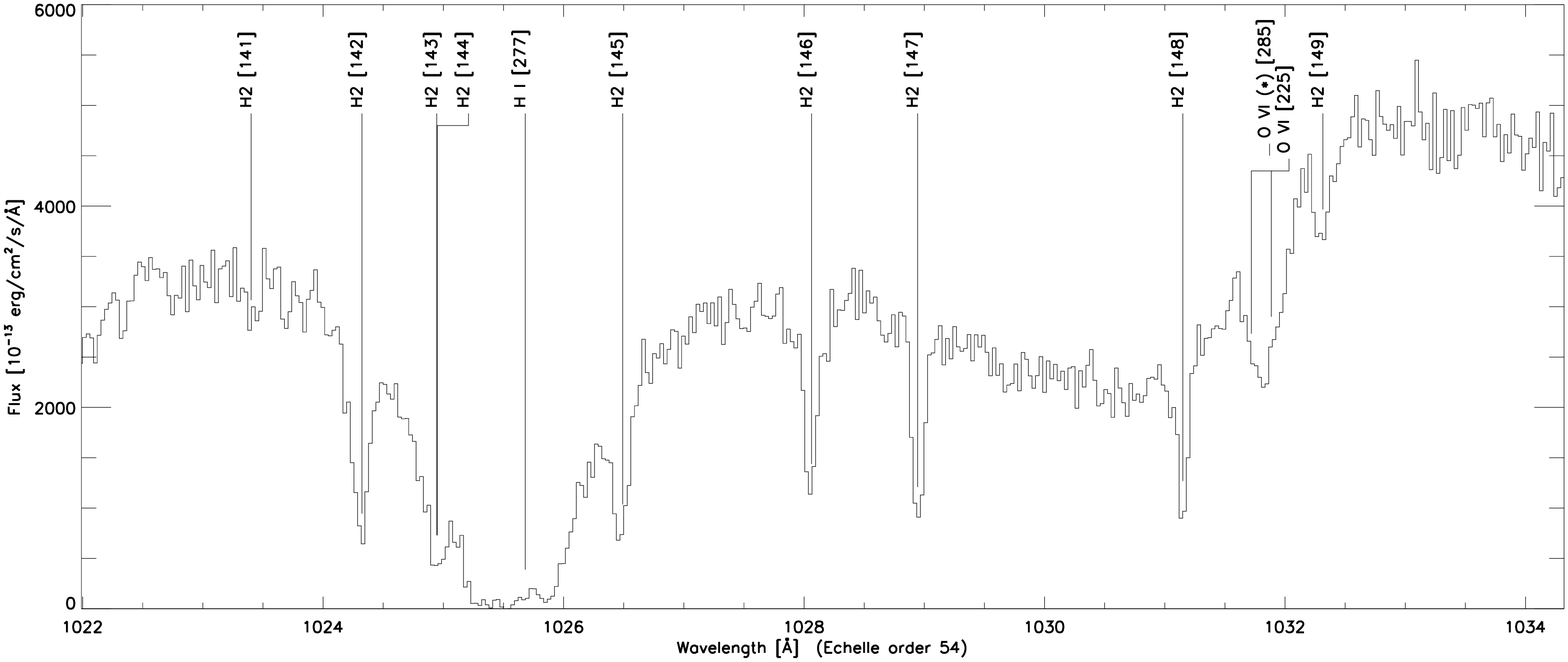}}
\resizebox{\hsize}{7.8cm}{\includegraphics{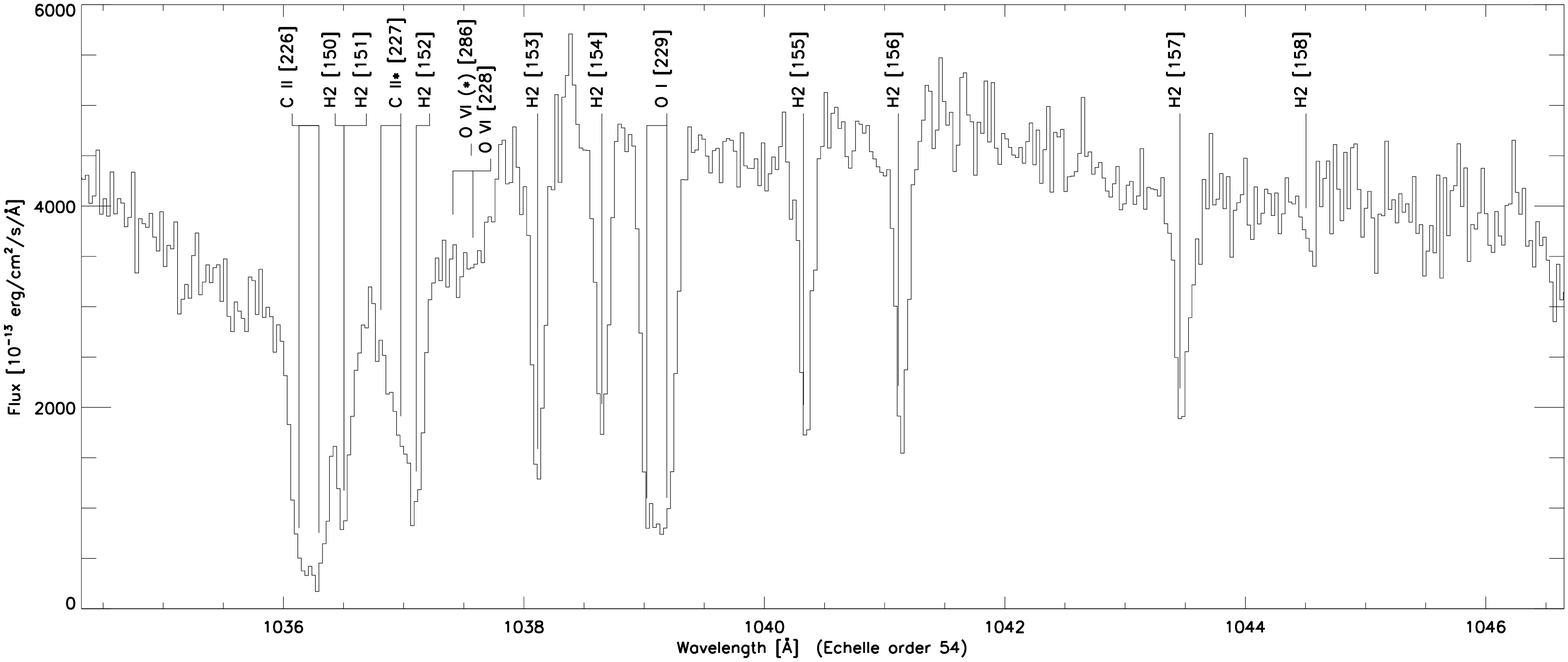}}
\end{figure*}
\begin{figure*}
\resizebox{\hsize}{7.8cm}{\includegraphics{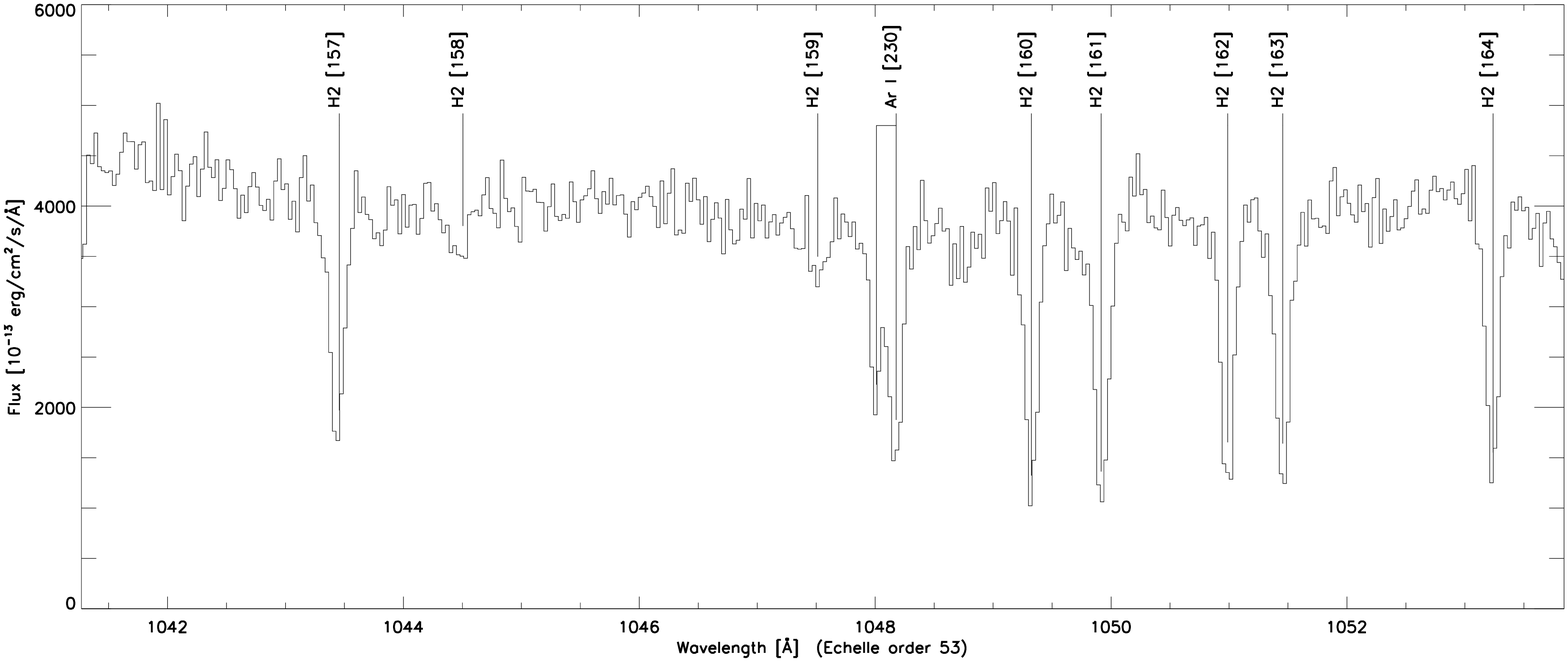}}
\resizebox{\hsize}{7.8cm}{\includegraphics{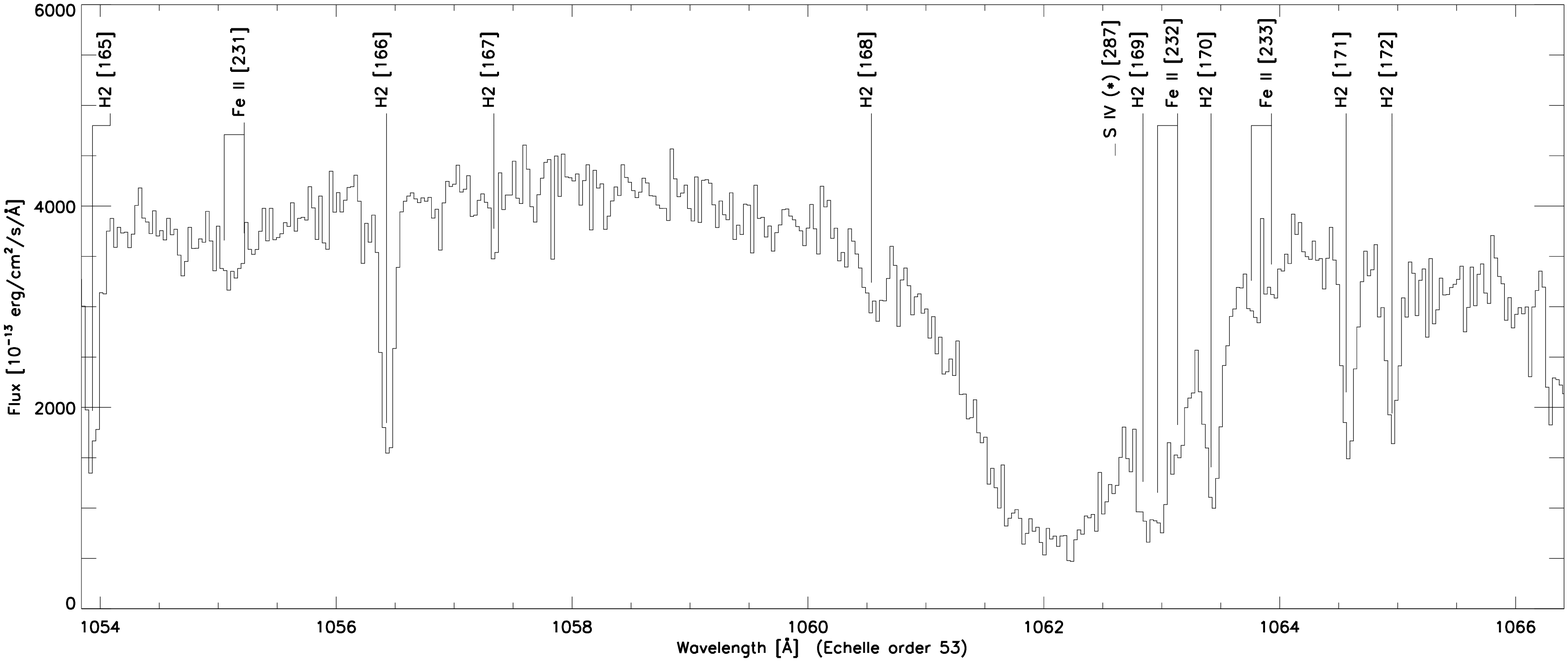}}
\resizebox{\hsize}{7.8cm}{\includegraphics{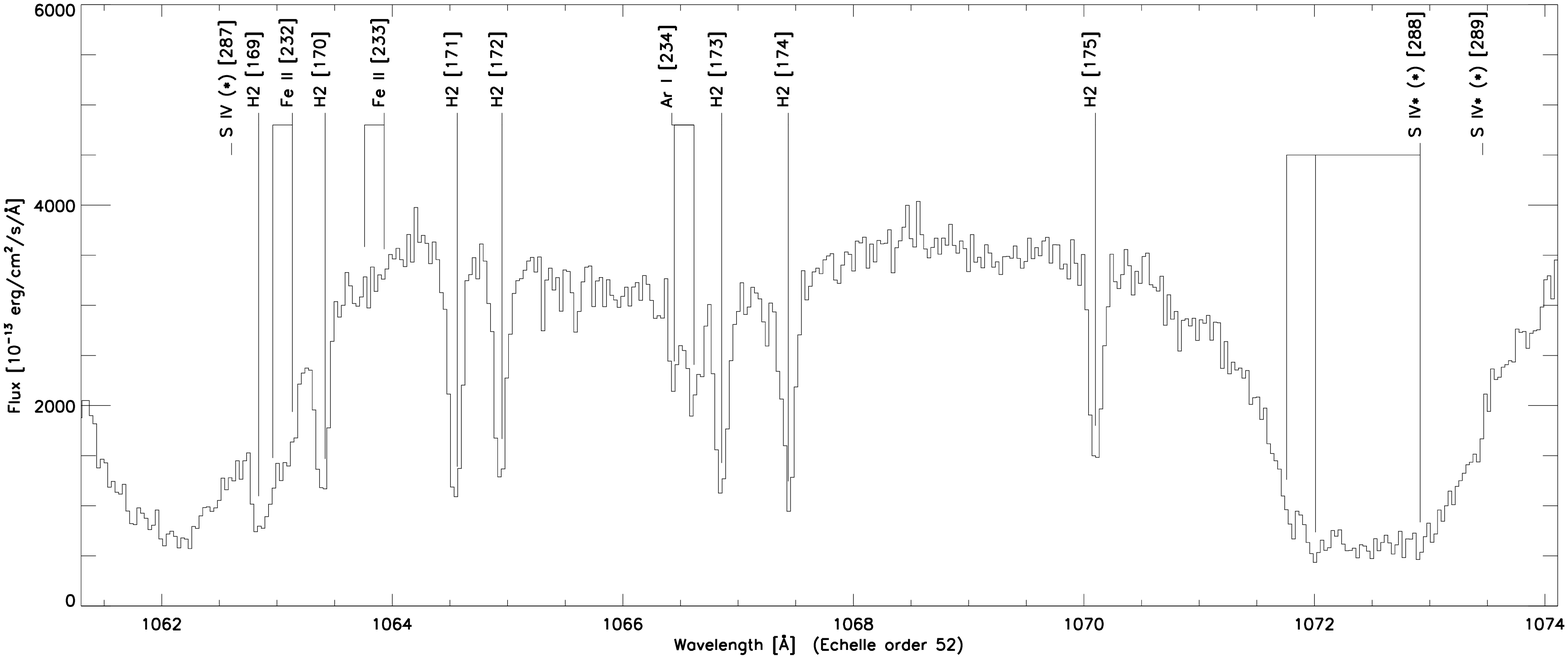}}
\end{figure*}
\begin{figure*}
\resizebox{\hsize}{7.8cm}{\includegraphics{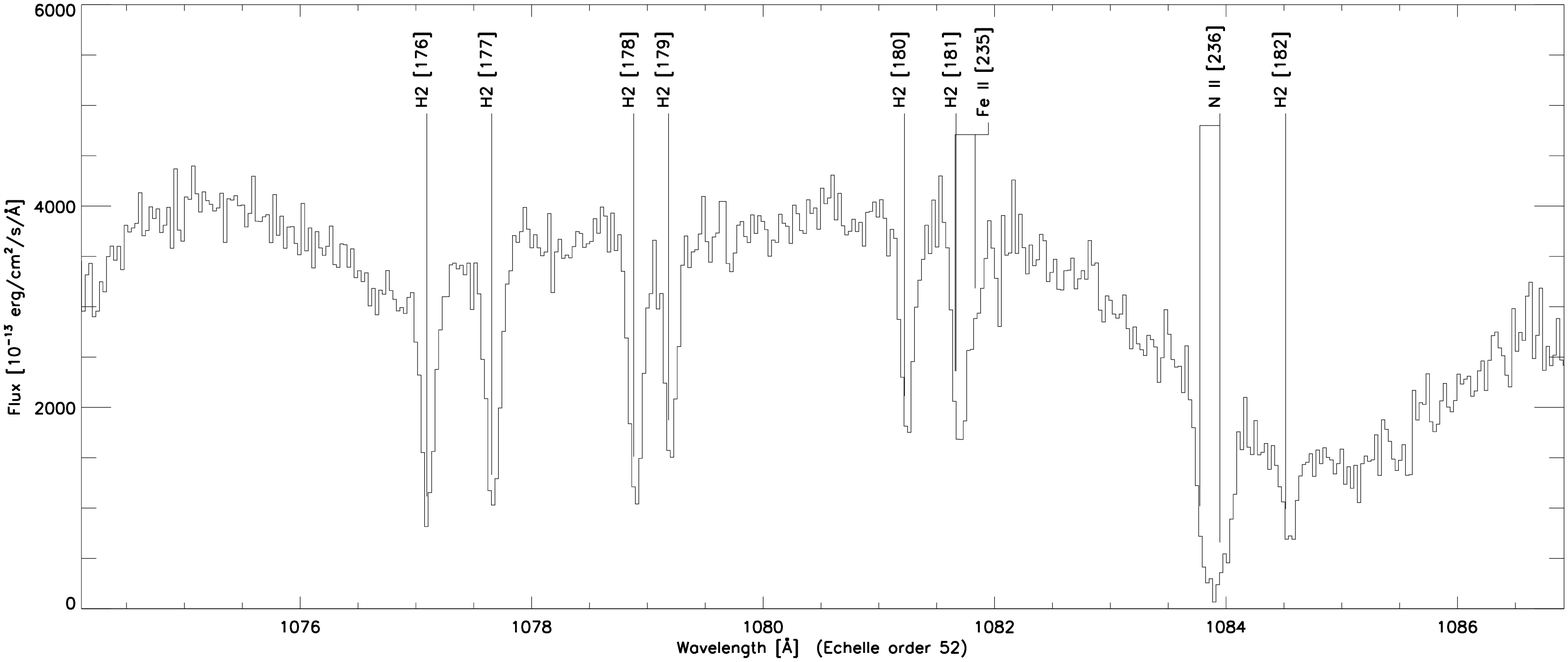}}
\resizebox{\hsize}{7.8cm}{\includegraphics{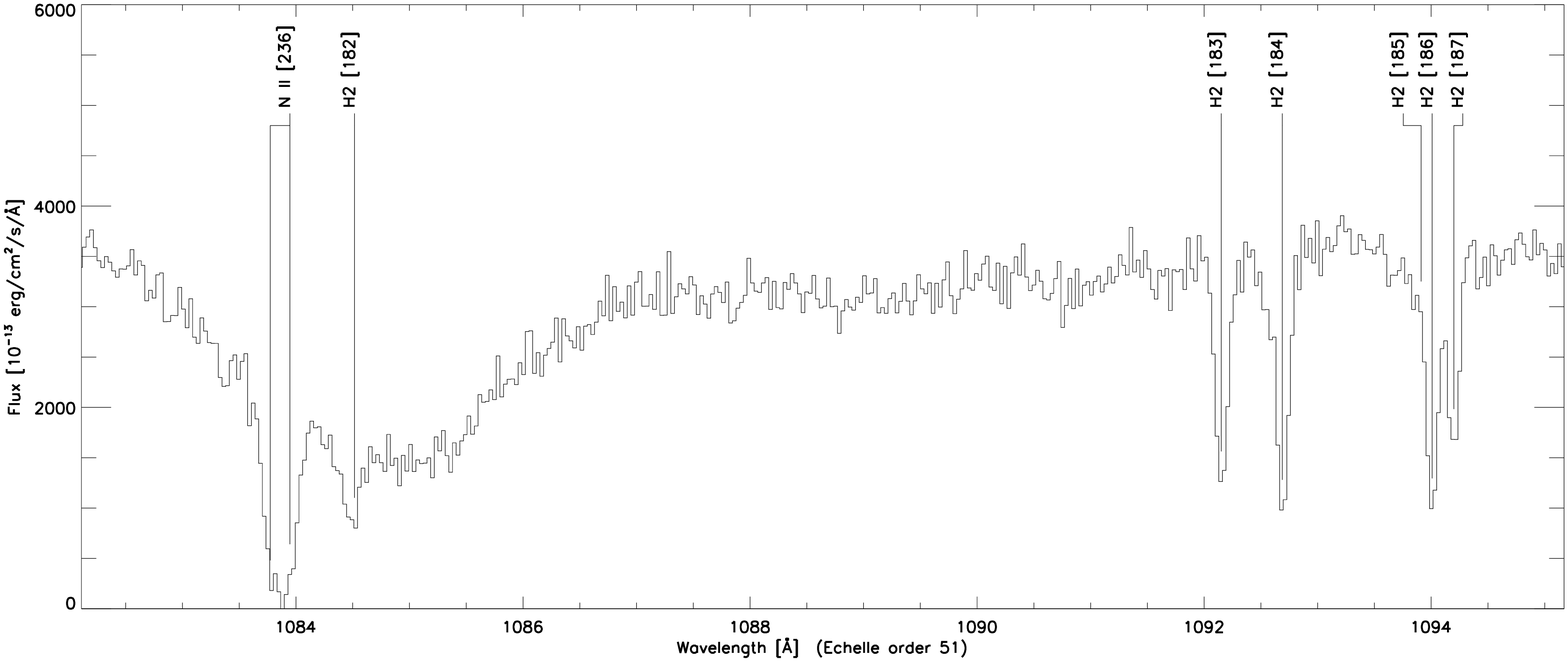}}
\resizebox{\hsize}{7.8cm}{\includegraphics{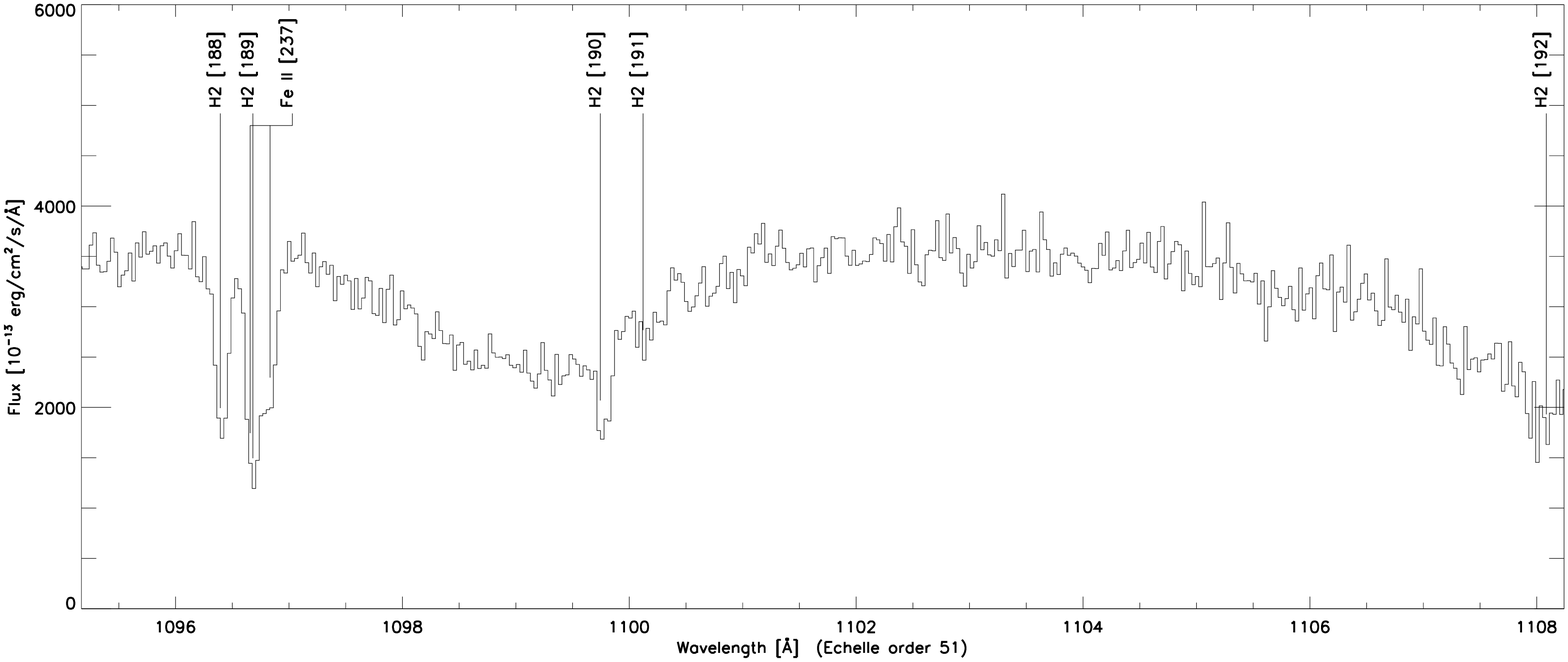}}
\end{figure*}
\begin{figure*}
\resizebox{\hsize}{7.8cm}{\includegraphics{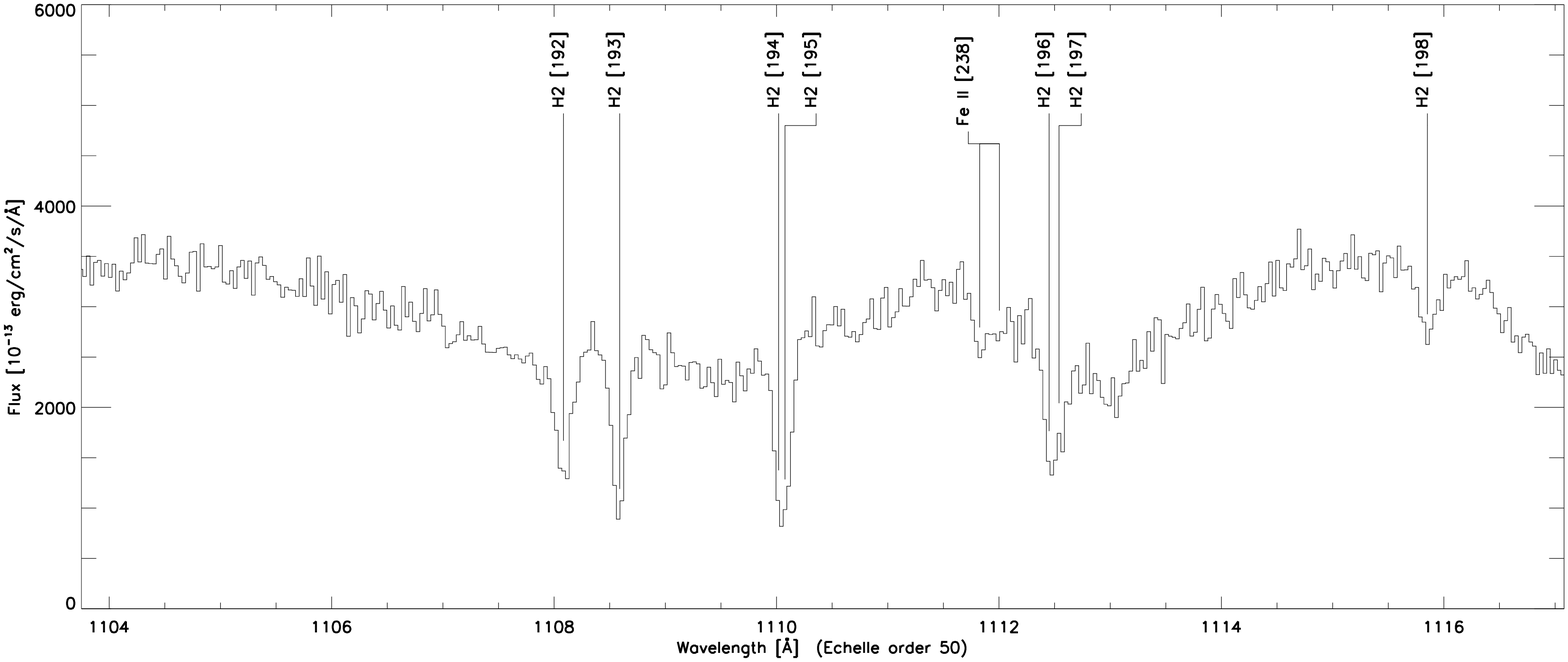}}
\resizebox{\hsize}{7.8cm}{\includegraphics{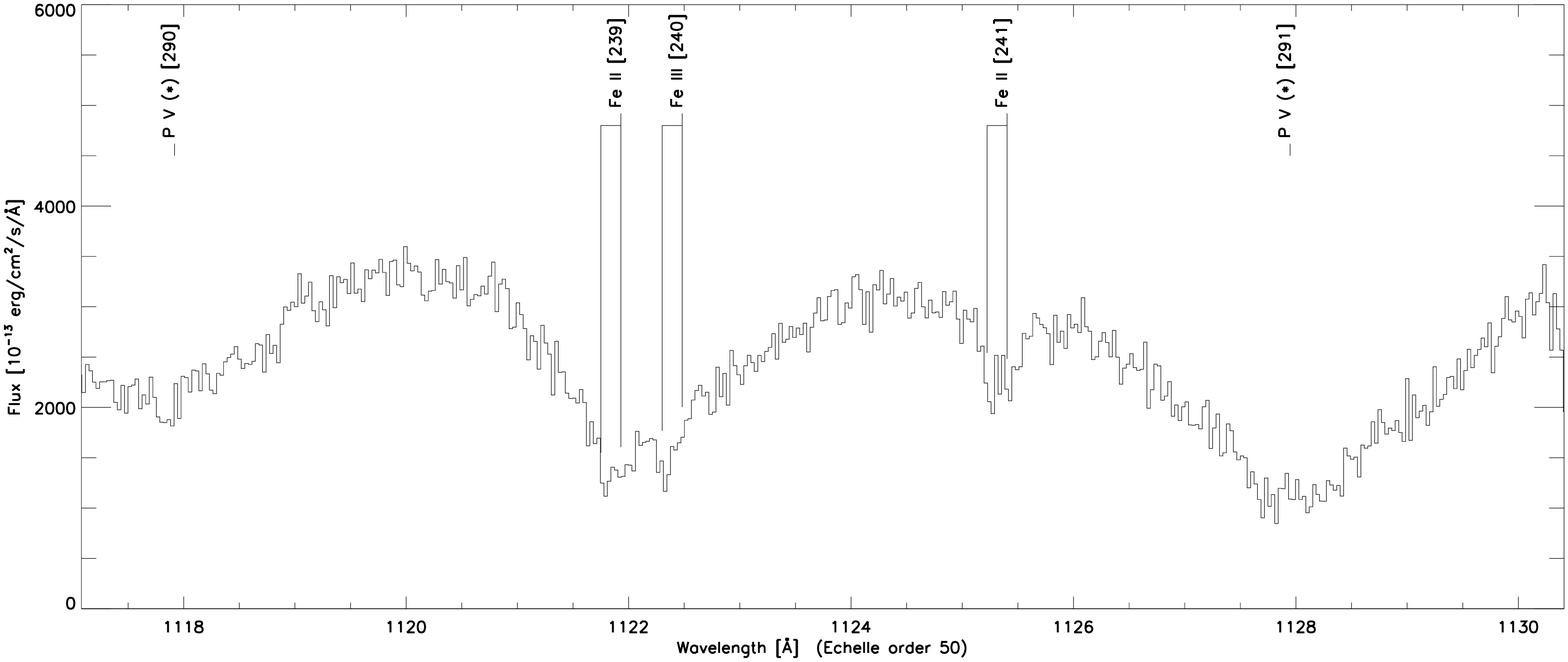}}
\resizebox{\hsize}{7.8cm}{\includegraphics{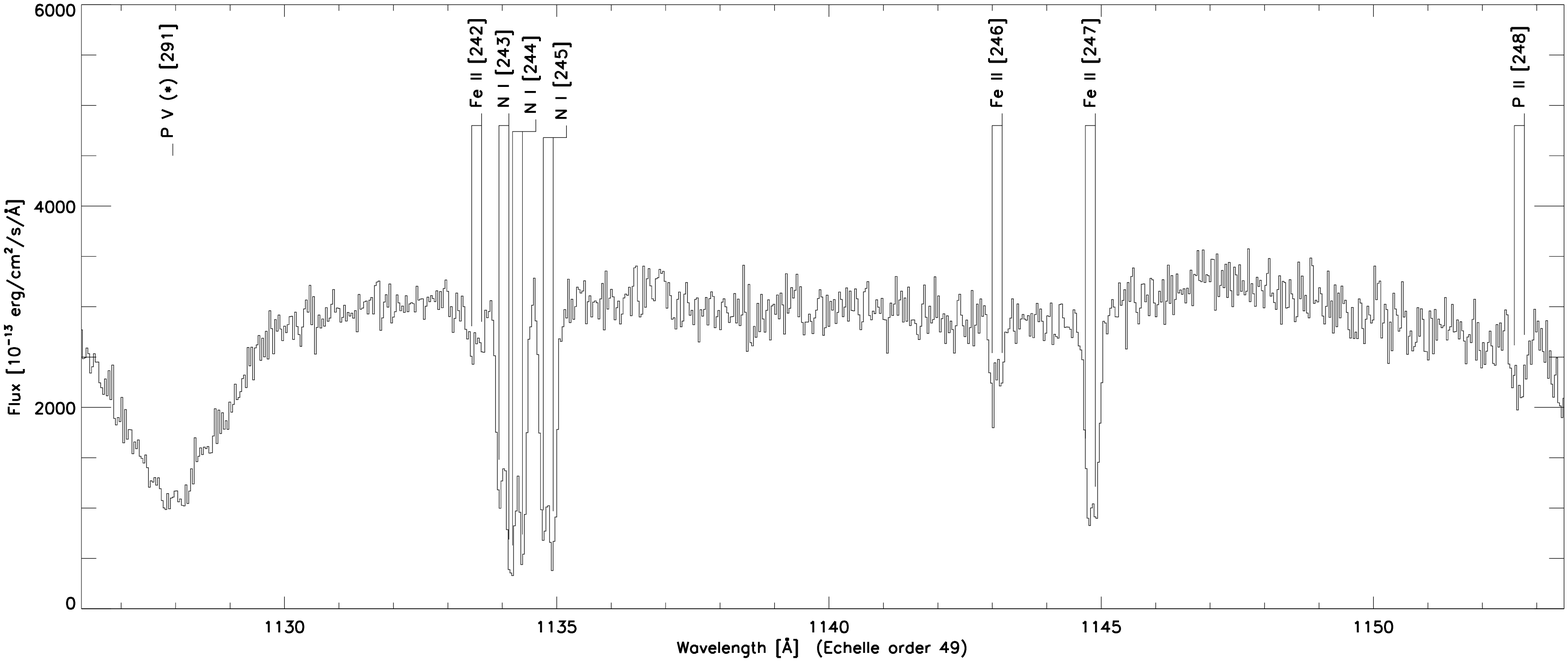}}
\end{figure*}
\begin{figure*}
\resizebox{\hsize}{7.8cm}{\includegraphics{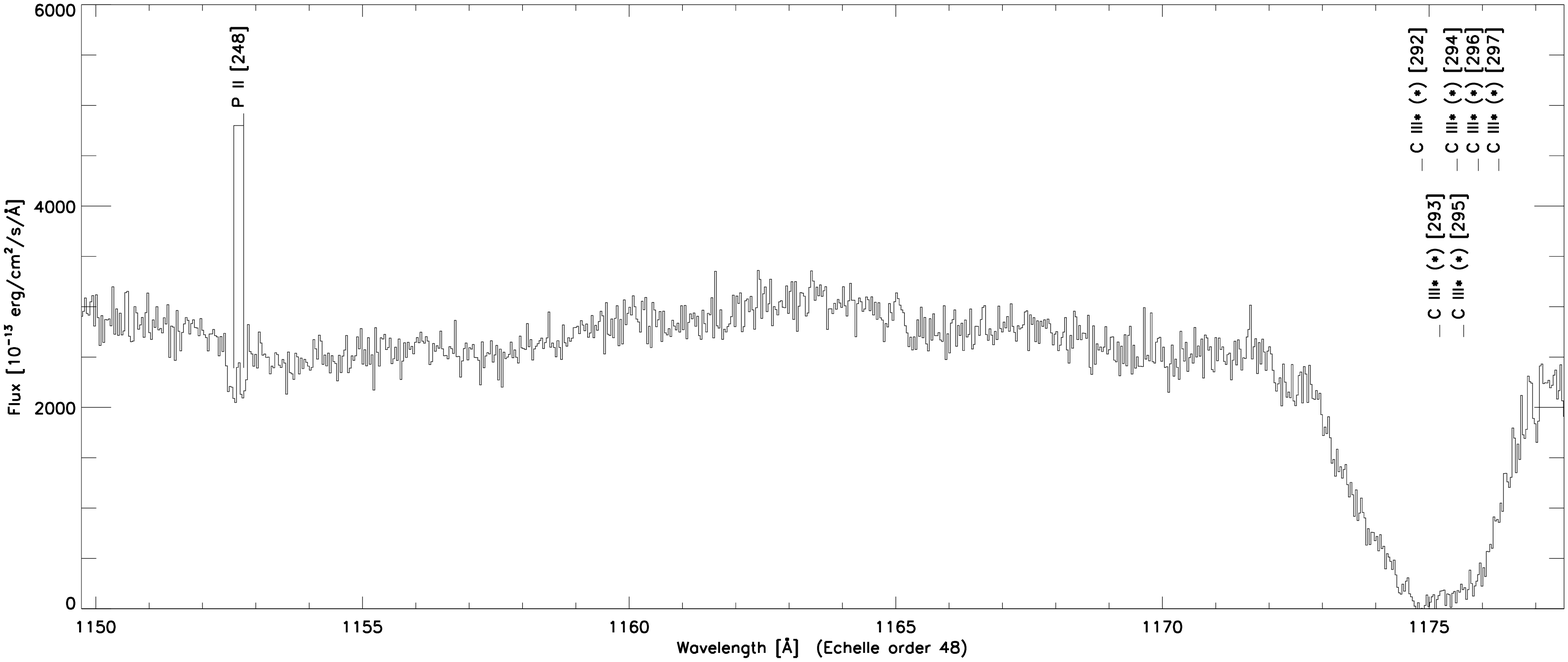}}
\resizebox{\hsize}{7.8cm}{\includegraphics{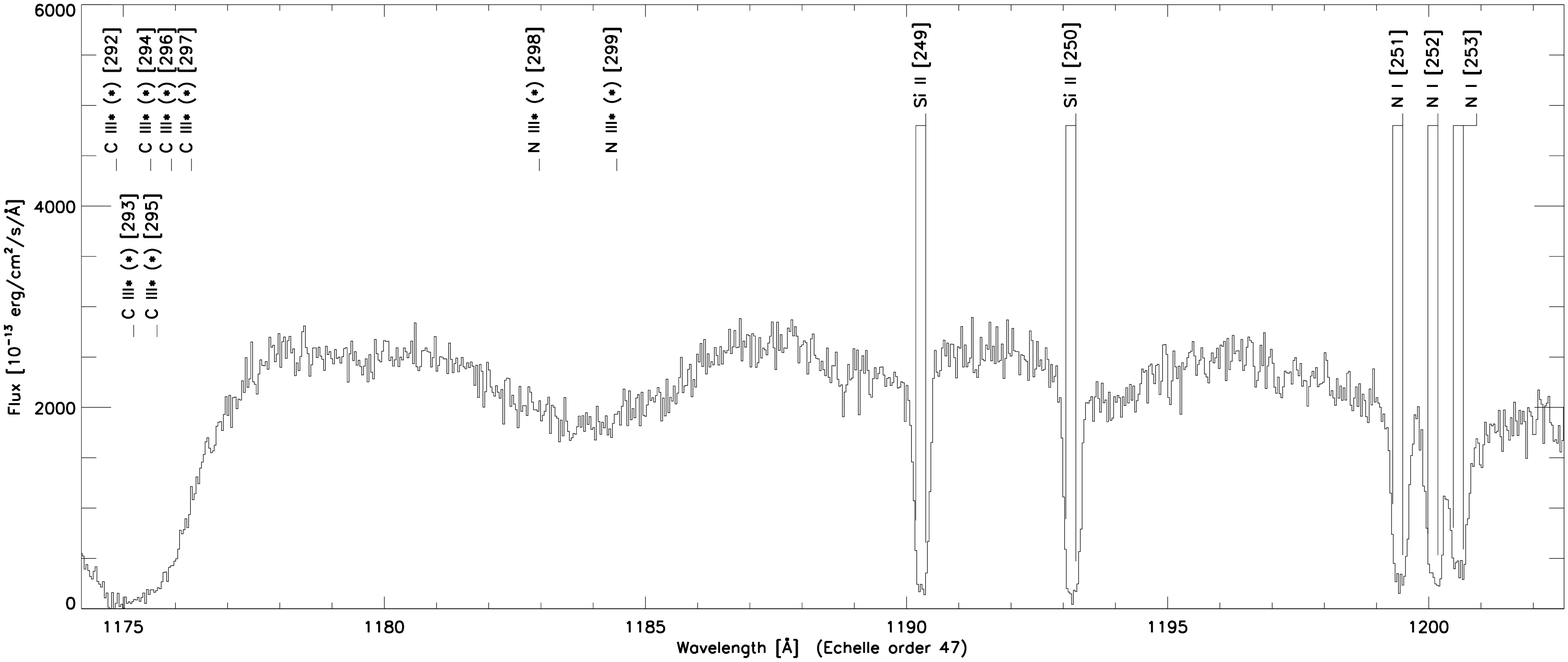}}
\resizebox{\hsize}{7.8cm}{\includegraphics{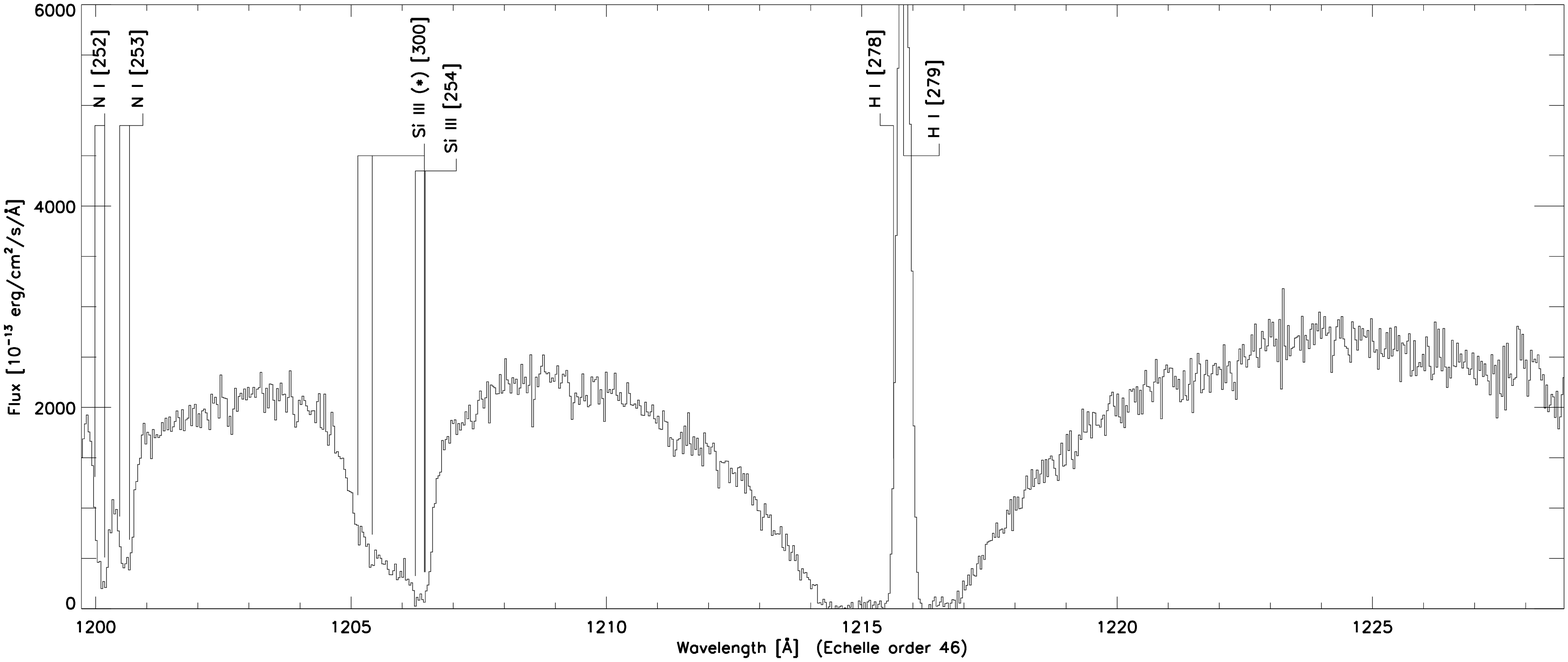}}
\end{figure*}
\begin{figure*}
\resizebox{\hsize}{7.8cm}{\includegraphics{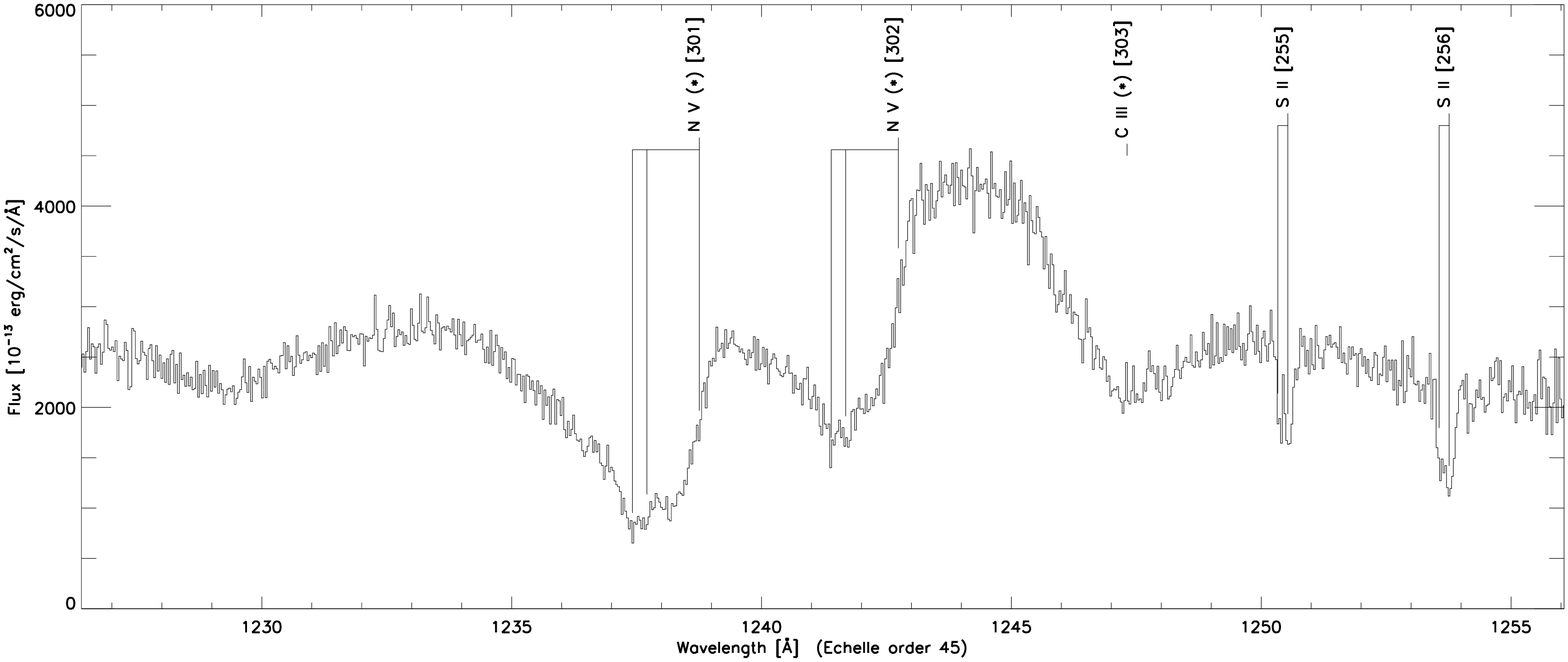}}
\resizebox{\hsize}{7.8cm}{\includegraphics{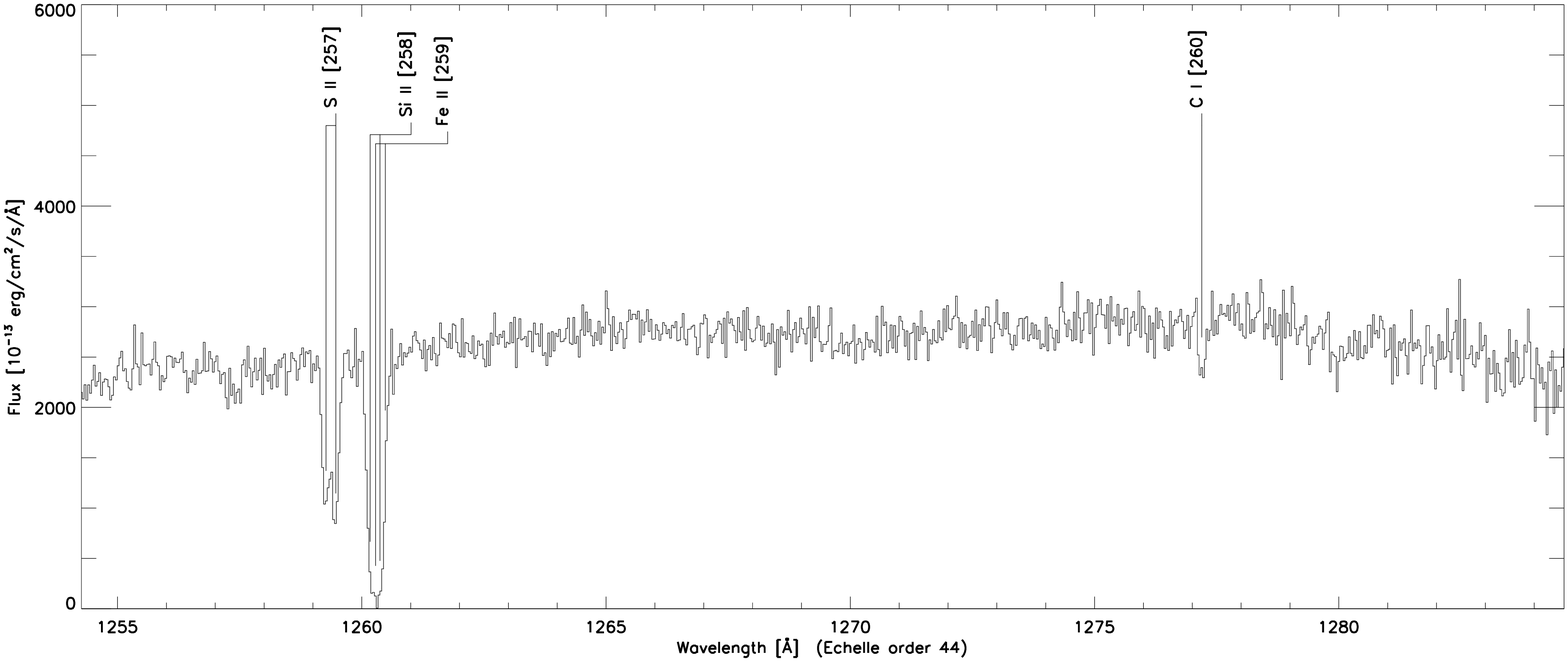}}
\resizebox{\hsize}{7.8cm}{\includegraphics{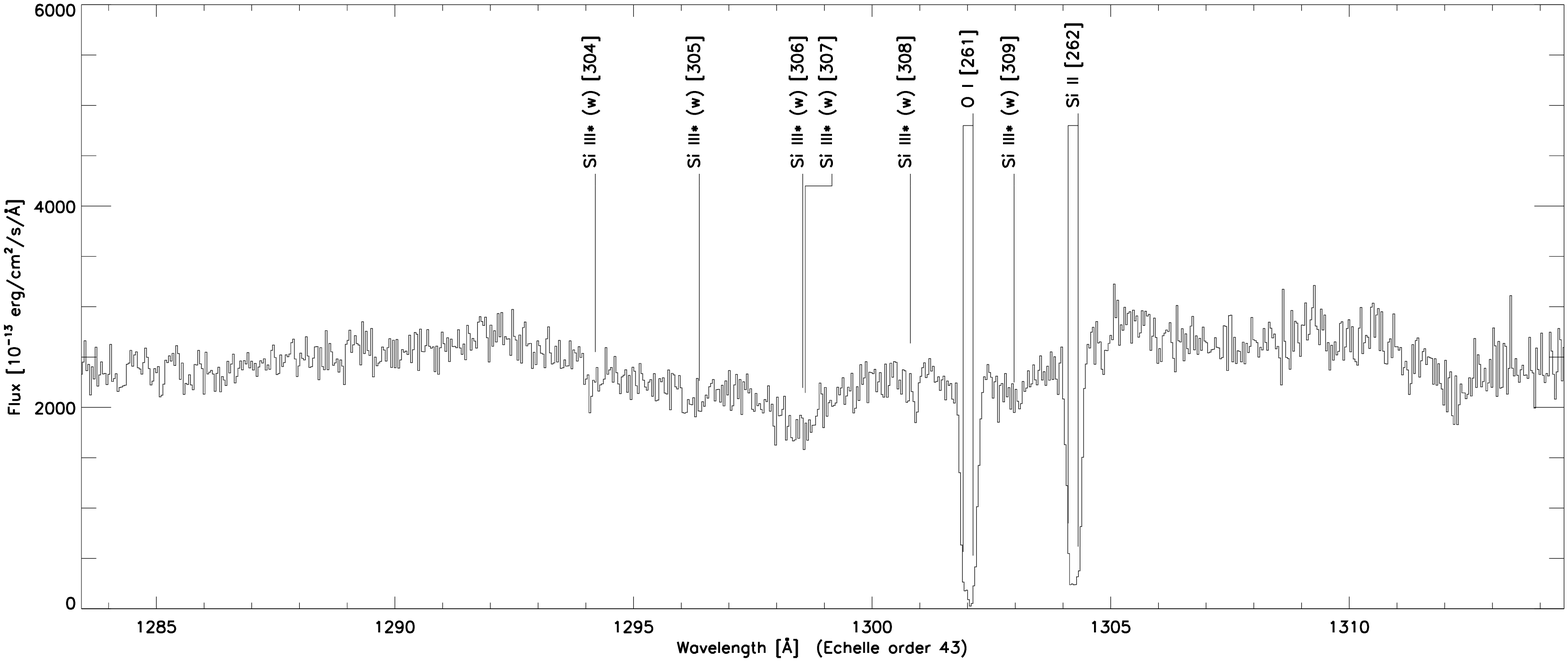}}
\end{figure*}
\begin{figure*}
\resizebox{\hsize}{7.8cm}{\includegraphics{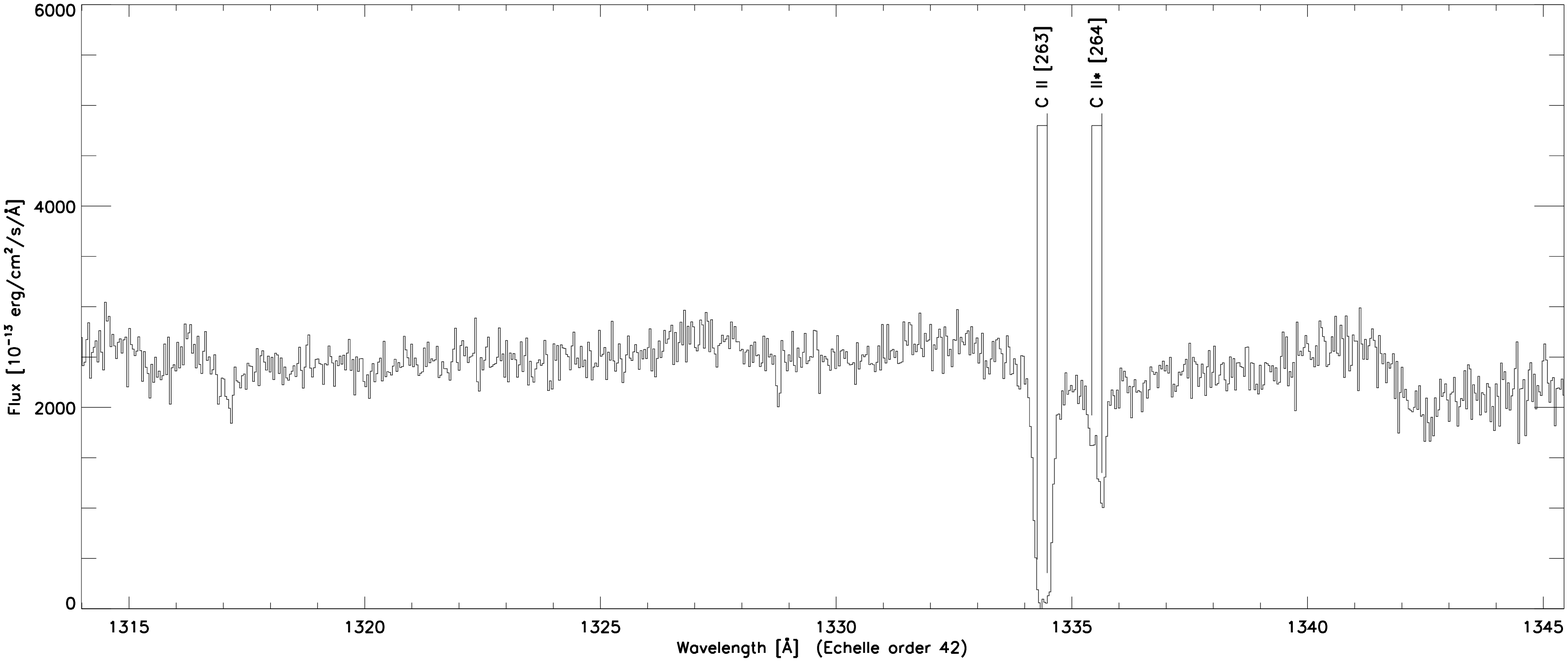}}
\resizebox{\hsize}{7.8cm}{\includegraphics{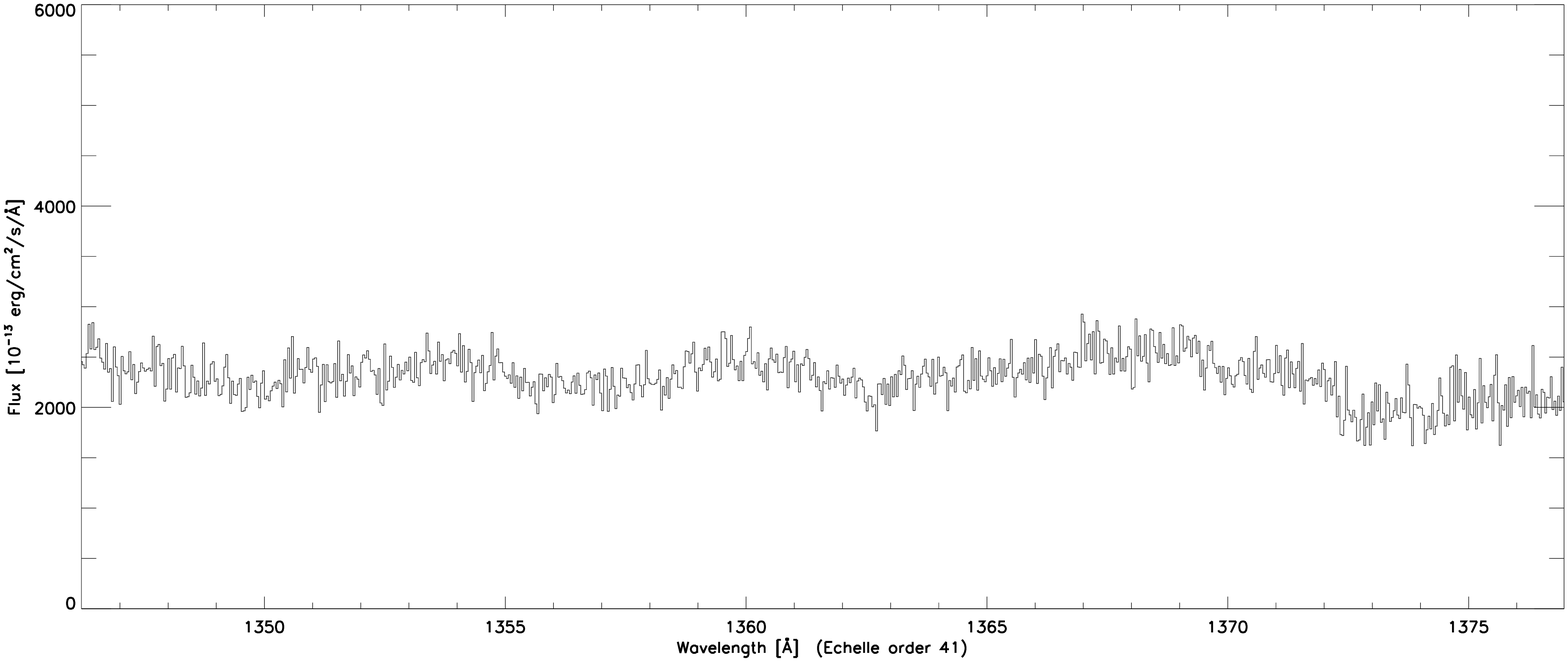}}
\resizebox{\hsize}{7.8cm}{\includegraphics{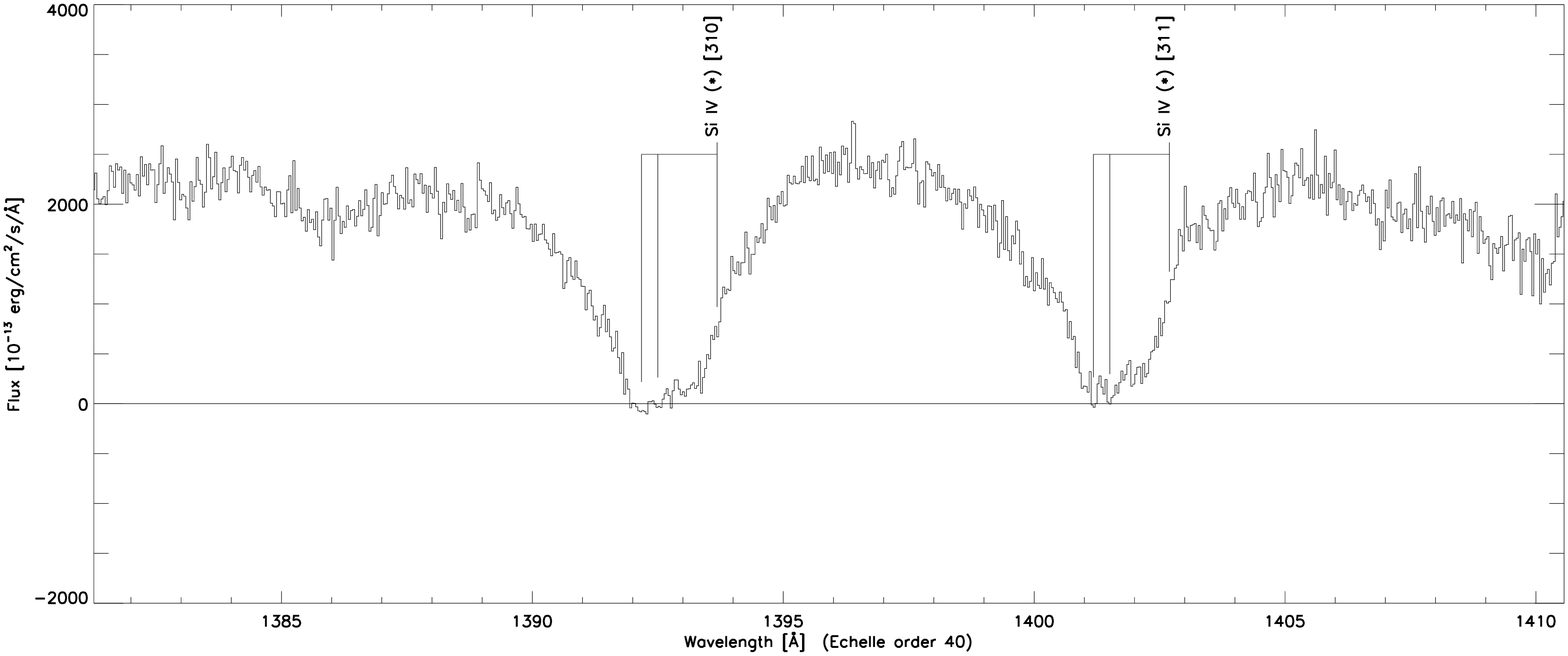}}
\end{figure*}

%\listofobjects
\end{document}